\documentclass[journal]{IEEEtran}

\ifCLASSOPTIONcompsoc
  \usepackage[nocompress]{cite}
\else
  \usepackage{cite}
\fi
\ifCLASSINFOpdf
\else
\fi
\usepackage{subfigure,balance}
\usepackage{color}
\def \etal {et al.}
\usepackage{booktabs}
\usepackage{multirow} 
\usepackage{mathtools}
\usepackage{enumitem}
\usepackage{cite}
\usepackage{url}
\usepackage{svg}
\usepackage{amsmath,amsfonts}
\usepackage{hyperref}
\usepackage{pifont}
\newcommand{\xmark}{\ding{55}}

\usepackage{graphicx}    
\usepackage{array}       
\usepackage{booktabs}    

\begin{document}
%
\title{The MSP-Podcast Corpus}
%
%
%
%

\author{Carlos~Busso,~\IEEEmembership{Fellow,~IEEE,}
        Reza~Lotfian,
        Kusha~Sridhar,
        Ali~N.~Salman,
        Wei-Cheng~Lin,~\IEEEmembership{Member,~IEEE,}
        Lucas~Goncalves,~\IEEEmembership{Member,~IEEE,}   
        Srinivas~Parthasarathy,~\IEEEmembership{Member,~IEEE,} 
        Abinay~Reddy~Naini,~\IEEEmembership{Student~Member,~IEEE,}
        Seong-Gyun~Leem,
        Luz~Martinez-Lucas~\IEEEmembership{Student-Member,~IEEE,}              
        Huang-Cheng~Chou,~\IEEEmembership{Member,~IEEE}
        and~Pravin Mote,~\IEEEmembership{Student~Member,~IEEE}
\IEEEcompsocitemizethanks{
\IEEEcompsocthanksitem This work was supported by the National Science Foundation (NSF) under
Grants CNS-1823166, CNS-2016719 and CAREER IIS-1453781. 
\IEEEcompsocthanksitem C. Busso is with the Language Technologies Institute, Carnegie Mellon University, Pittsburgh PA-15213 USA (busso@cmu.edu).
\IEEEcompsocthanksitem R. Lotfian is with Athenahealth, Boston, MA, USA (rlotfian@athenahealth.com).
\IEEEcompsocthanksitem K. Sridhar is with Accenture LLP, Mountain View, CA, USA (k.sridhara.murthy@accenture.com).
\IEEEcompsocthanksitem A. Salman is with ARRAY Innovation, Bahrain (ali.salman@array.world).
\IEEEcompsocthanksitem W.-C. Lin is with Bosch Center for Artificial Intelligence, Bosch Research, Pittsburgh PA-15222 USA (wei-cheng.lin@us.bosch.com).
\IEEEcompsocthanksitem L. Goncalves and S. Parthasarathy are with Amazon, USA (sglucas@amazon.com).
\IEEEcompsocthanksitem A. Reddy~Naini, L. Martinez-Lucas and P. Mote are with the Erik Jonsson School of Engineering and Computer Science, University of Texas at Dallas, Richardson, TX 75080 USA(AbinayReddy.Naini@utdallas.edu, luz.martinez-lucas@utdallas.edu, Pravin.Mote@UTDallas.edu).
\IEEEcompsocthanksitem S.-G. Leem is with the Reality Labs at Meta Platforms, Inc. (sgleem@meta.com).
\IEEEcompsocthanksitem H.-C. Chou is with the Signal Analysis and Interpretation Laboratory (SAIL), Ming Hsieh Department of Electrical and Computer Engineering, University of Southern California (USC), Los Angeles, CA 90089, USA (huangchengchou@gmail.com).}
\thanks{Manuscript received September 10, 2025; revised ?}}

\IEEEtitleabstractindextext{%
\begin{abstract}
The availability of large, high-quality emotional speech databases is essential for advancing \emph{speech emotion recognition} (SER) in real-world scenarios. However, many existing databases face limitations in size, emotional balance, and speaker diversity. This study describes the MSP-Podcast corpus, summarizing our ten-year effort. The corpus consists of over 400 hours of diverse audio samples from various audio-sharing websites, all of which have Common Licenses that permit the distribution of the corpus. We annotate the corpus with rich emotional labels, including primary (single dominant emotion) and secondary (multiple emotions perceived in the audio) emotional categories, as well as emotional attributes for valence, arousal, and dominance. At least five raters annotate these emotional labels. The corpus also has speaker identification for most samples, and human transcriptions of the lexical content of the sentences for the entire corpus. The data collection protocol includes a machine learning-driven pipeline for selecting emotionally diverse recordings, ensuring a balanced and varied representation of emotions across speakers and environments. The resulting database provides a comprehensive, high-quality resource, better suited for advancing SER systems in practical, real-world scenarios.
\end{abstract}

}

\maketitle

\IEEEdisplaynontitleabstractindextext

%
\IEEEpeerreviewmaketitle

\ifCLASSOPTIONcompsoc
\IEEEraisesectionheading{\section{Introduction}\label{sec:introduction}}
\else
\section{Introduction}
\label{sec:introduction}
\fi

%
%
%
%

\IEEEPARstart{A}{ffective} computing is a prominent research field focused on understanding, analyzing, recognizing, and synthesizing human emotions. Enriching interfaces with emotional awareness has the potential to enable significant applications across diverse domains, including \emph{human-computer interaction} (HCI), mental health, security and defense, education, and entertainment. Among the various modalities, speech plays a critical role in these interfaces by conveying information beyond the literal meaning of words. However, recognizing emotions from speech in realistic settings poses considerable challenges, largely due to the subtle and complex expressive behaviors inherent in human interactions \cite{Busso_2013}. To effectively develop and evaluate methods that address naturalistic scenarios, it is crucial to have access to datasets that accurately represent these real-world conditions. A common issue in building \emph{speech emotion recognition} (SER) systems is the limited availability of datasets that provide sufficient data, diversity, and representativeness of naturalistic interactions. This scarcity impedes further advancements in the field of speech affective computing and related research areas.

Over the years, numerous studies have focused on developing diverse methods for collecting emotionally rich databases. These approaches include using actors delivering predefined sentences with specific emotional states \cite{Cao_2014_2, Liberman_2002, Burkhardt_2005, Livingstone_2018}, employing speakers in semi-structured scenarios designed to evoke natural emotional responses \cite{Busso_2008_5,Douglas-Cowie_2000,Busso_2017}, recording colloquial conversation between participants  \cite{Ringeval_2013,Chou_2017}, utilizing acted TV shows as source for emotional content \cite{Shen_2020,Grimm_2008, Poria_2019}, and collecting data from audio and video sharing platforms \cite{Vidal_2020, Upadhyay_2023_2,Kollias_2019, Lotfian_2019_3,Martinez-Lucas_2020}. However, utilizing some of these aforementioned methods comes with issues. Using actors with predefined sentences often results in exaggerated or stereotypical emotional expressions that may not reflect natural human behavior. The scripted nature also limits variability and spontaneity, potentially biasing models trained on such datasets. Semi-structured scenarios aim for more spontaneity, but may still fail to capture authentic emotional experiences. For example, the participants' awareness of being observed can influence their behavior, leading to unnatural responses. Acted TV shows, while providing large amounts of emotional material, face challenges such as exaggerated externalizations of emotions for dramatic effect and a lack of authenticity. Additionally, the context in TV shows may not generalize well to real-world scenarios, and ethical and copyright issues can complicate the use of such data in research. These limitations highlight the need for the MSP-Podcast corpus, which contains naturalistic, diverse, and well-annotated data to advance the study of emotional states. Collecting authentic emotional data in real-world settings without scripts or actors can provide more genuine samples. The diversity in the data is crucial for developing models that generalize across various scenarios and contexts. Moreover, including a variety of annotator opinions ensures that the dataset can more accurately capture the complexity of human emotions.

This paper presents the MSP-Podcast corpus, summarizing our 10-year effort to collect this corpus. Mariooryad \etal \cite{Mariooryad_2014_3} presented the initial idea for a scalable data collection protocol that inspired our effort for the MSP-Podcast corpus. Lotfian and Busso \cite{Lotfian_2019_3} formulated a protocol for using machine-learning methods to retrieve emotional recordings that are carefully annotated with emotional labels. The focus of this paper is to describe the resulting database, detailing the changes made to the protocol to enhance the quality of the data. The final release of the MSP-Podcast corpus comprises 409 hours of annotated data collected from more than 3,641 speakers, incorporating diverse audio samples from various sources with diverse emotional content. 
The continuous growth of multimedia content on the Internet offers an abundant resource for audio data, particularly podcasts that cover a wide array of topics and emotions. Our primary challenge was to select audio segments that provide a balanced representation across the emotional spectrum. We carefully selected and downloaded podcasts featuring natural conversations among various speakers on diverse subjects, including both positive and negative topics, such as personal stories, debates, and cultural discussions. To ensure the database can be shared widely within the research community, we focused on recordings available under Creative Commons licenses with minimal restrictions. The audio was processed to extract clean, single-speaker segments by removing silence, background noise, music, and overlapping speech, utilizing advanced algorithms for voice activity detection, speaker diarization, and noise estimation. We employed enhanced machine learning models trained on larger corpora to identify segments exhibiting specific emotional categories and values for the attributes of valence (negative versus positive), arousal (calm versus active), and dominance (weak versus strong). This refined approach enables greater control over the emotional content, increases speaker diversity, and preserves the spontaneous nature of the recordings.

This paper presents our methods for curating a more diverse and emotionally rich set of naturalistic speech samples from podcasts available on audio-sharing platforms. We describe the emotional annotation process, which began with crowdsourcing evaluations and continued with a carefully controlled annotation process involving trained students from our institution. At least five raters annotated each speaking turn, providing rich labels for primary (single dominant emotion) and secondary (all emotions perceived in the speech) emotional categories, and emotional attributes for valence, arousal, and dominance. We describe our strategy to enhance the quality of annotations, which includes tracking the performance of annotators on a weekly basis, providing detailed feedback, and implementing a training strategy to improve their annotations if their quality falls below a given threshold. We also describe other annotations included in the corpus, including speaker identification for most of the corpus and human transcriptions, with a focus on the quality control methods we implemented. The contribution of this study is not only the resulting database but also the lessons learned from this multi-year effort, which can guide future data collections.

The remainder of this paper is organized as follows. Section \ref{sec:related} provides a brief overview of existing emotional databases. Section \ref{sec:Protocol} outlines the protocol used for data collection, including the selection of podcasts, segmentation into short turns, post-processing and filtering steps, and procedures for emotional annotation. Section \ref{sec:annotation} describes the annotations of the corpus, including emotions, speaker information, and lexical content. Section \ref{sec:Organization} provides the partitions of the corpus and a brief recollection of early releases of this corpus. Section \ref{sec:baseline} presents SER baselines for classifying primary emotions and predicting emotional attributes. Section \ref{sec:discussion} highlights new research opportunities opened by key features of this corpus. Finally, Section \ref{sec:conclusion} concludes the paper with a summary and final remarks.



\section{Related Work}
\label{sec:related}

\subsection{Emotional Databases}
\label{ssec:databases}

Table \ref{tab:corpus_table} presents some emotional databases. Although the research community has access to numerous emotional databases, they come with certain limitations that restrict their effectiveness in tackling ongoing research problems. These limitations include the lack of naturalness in the emotional expressions, unbalanced emotional content, and constraints in size and speaker diversity.

\begin{table*}[h!]
\centering
\caption{Overview of Speech Emotion Databases}
\label{tab:corpus_table}
\begin{tabular}{@{}lccccllc@{}}
\toprule
\textbf{Corpus} & \textbf{Size} & \textbf{\#spk} & \textbf{Avail} & \textbf{Size} & \textbf{\# Spkr} & \textbf{Type} & \textbf{Lang.} \\
\midrule
MSP-PODCAST 2.0 (this paper) & \checkmark & \checkmark & \checkmark & 407h & xxx & Spontaneous & English \\
Dusha \cite{Kondratenko_2023} & \checkmark & \checkmark & \checkmark & 346h36m & 8,308 & Acted, Spontaneous & Russian \\
Crowdsourcing Emotional Speech \cite{Smith_2018} & \checkmark & \checkmark & \checkmark & 187h & 2,965 & Spontaneous & English \\
BIIC-Podcast \cite{Upadhyay_2023_2} & \checkmark & \xmark & \checkmark & 147h26m & Unknown & Spontaneous & Taiwanese Mandarin \\
MIKU-EmoBench \cite{Cheng_2025} & \checkmark & \xmark & \checkmark & 131h12m & Unknown & Spontaneous & Multiple \\
CMU-MOSEAS \cite{Zadeh_2020} & \checkmark & \checkmark & \checkmark & 68h49m & 1,645 & Spontaneous & Multiple \\
CMU-MOSEI \cite{Zadeh_2018} & \checkmark & \checkmark & \checkmark & 65h53m & 1,000 & Spontaneous & English \\
THAI-SER\cite{Wongpithayadisai_2025} & \xmark & \checkmark & \checkmark & 41h36m & 200 & Acted & Thai \\
CEMO\cite{Devillers_2006} & \xmark & \checkmark & \checkmark & 20h & 688 & Spontaneous & French \\
IEMOCAP\cite{Busso_2008_5} & \xmark & \xmark & \checkmark & 12h26m & 10 & Acted & English \\
MELD\cite{Poria_2019} & \xmark & \checkmark & \checkmark & 30h45m & 407 & Acted & English \\
TUM AVIC\cite{Schuller_2007_3} & \xmark & \xmark & \checkmark & 10h23m & 21 & Spontaneous & English \\
MSP-IMPROV\cite{Busso_2017} & \xmark & \xmark & \checkmark & 9h35m & 12 & Acted & English \\
FAU-AIBO\cite{Batliner_2008} & \xmark & \xmark & \checkmark & 9h12m & 51 & Spontaneous & German \\
CHEAVD 2.0 \cite{Li_2018_5} & \xmark & \checkmark & \checkmark & 7h54m & 527 & Acted & Mandarin \\
DEMoS\cite{parada_2020} & \xmark & \xmark & \checkmark & 7h40m & 68 & Induced & Italian \\
Emozionalmente v1.1\cite{Catania_2025} & \xmark & \checkmark & \checkmark & 7h18m & 431 & Acted & Italian \\
WHiSER\cite{Naini_2024_2} & \xmark & \xmark & \checkmark & 6h21m & Unknown & Spontaneous & English \\
SEMAINE\cite{McKeown_2012} & \xmark & \xmark & \checkmark & 6h30m & 20 & Induced & English \\
Chen Bimodal\cite{Chen_2000_2} & \xmark & \checkmark & \xmark & 5h36m & 100 & Acted & English \\
CREMA-D\cite{Cao_2014_2} & \xmark & \xmark & \checkmark & 5h16m & 91 & Acted & English \\
NNIME\cite{Chou_2017} & \xmark & \checkmark & \checkmark & 11h & 43 & Acted & Taiwanese Mandarin \\
UrduSER\cite{Akhtar_2025} & \xmark & \xmark & \checkmark & 3h2m & 10 & Acted & Urdu \\
RECOLA\cite{Ringeval_2013} & \xmark & \xmark & \checkmark & 3h50m & 46 & Spontaneous & French \\
CMU-MOSI\cite{Zadeh_2016} & \xmark & \xmark & \checkmark & 2h34m & 98 & Spontaneous & English \\
VAM-Audio\cite{Grimm_2008} & \xmark & \xmark & \checkmark & 48m & 47 & Spontaneous & German \\
Emo-DB\cite{Burkhardt_2005} & \xmark & \xmark & \checkmark & 3h & 10 & Acted & German \\
RAVDESS\cite{Livingstone_2018} & \xmark & \xmark & \checkmark & 7,356 samples & 24 & Acted & English \\
\bottomrule
\end{tabular}
\end{table*}

Traditional emotional corpora designed for emotion recognition largely depended on actors who were directed to vocalize sentences with intended emotions. This practice was used to create several well-known emotional databases, such as the Emo-DB \cite{Burkhardt_2005}, RAVDESS \cite{Livingstone_2018}, TESS \cite{pichora2020toronto}, CREMA-D \cite{Cao_2014_2}, and the Chen Bimodal \cite{Chen_2000_2} databases. While these datasets have played an essential role in early research efforts, the use of acted emotions presents challenges in truly mirroring the complex and spontaneous nature of genuine human emotions, as discussed by Devillers \etal \cite{Devillers_2005} and Batliner \etal \cite{Batliner_2000}. 
Some databases have been designed to address this limitation. The DUSHA corpus \cite{Kondratenko_2023} was constructed using a hybrid data collection methodology, combining elicited speech from non-professional actors with spontaneous speech extracted from podcasts. This approach aims to balance the experimental control inherent in acted performances with the ecological validity of naturalistic recordings.
Other databases, such as the USC-IEMOCAP \cite{Busso_2008}, MSP-IMPROV \cite{Busso_2017}, and THAI-SER \cite{Wongpithayadisai_2025} corpora, aimed to bridge this gap by incorporating more naturally occurring emotional expressions within dyadic interactions, thereby deviating from the more scripted monologues of previous databases. These endeavors made significant strides in producing dialogue that closely mimics the nuances of real-world emotional exchanges. Yet, the usage of professional actors remained a barrier to capturing naturalistic emotional responses.


In the pursuit of authenticity, other datasets have relied on spontaneous interactions derived from sources such as colloquial conversations (SEMAINE \cite{McKeown_2012}, RECOLA \cite{Ringeval_2013}, TUM-AVIC \cite{Schuller_2007_3}), television programs (VAM \cite{Grimm_2008}, MELD \cite{Poria_2019}, CHEAVD \cite{Li_2017_4}, UrduSER \cite{Akhtar_2025}), the Internet (BIIC-Podcast \cite{Upadhyay_2023_2}, WHiSER  \cite{Naini_2024_2}, CMU-MOSI \cite{Zadeh_2016}, CMU-MOSEI \cite{Zadeh_2018}, CMU-MOSEAS \cite{Zadeh_2020}), and customer service calls (CEMO \cite{Devillers_2006}). This shift towards spontaneity was critical in capturing genuine emotional displays, but these databases faced the obstacle of skewed emotional representations, constrained by the contexts from which they were sourced. For instance, television programs broadcasting relationship issues might lean towards negative emotions \cite{Grimm_2008}, while casual conversations might predominantly exhibit positive emotions \cite{Ringeval_2013}. The emotional imbalance also poses a challenge for SER models, which require diverse and evenly distributed emotional examples to learn effectively. For example, Naini \etal \cite{Naini_2025} demonstrated SER improvements by just undersampling the training set to match the emotional distribution of the target domain.


A prominent trend in emotion corpus development involves leveraging crowdsourcing to acquire data from a large pool of participants using their personal, consumer-grade devices. In this paradigm, exemplified by corpora such as Emozionalmente \cite{Catania_2025} and the dataset by Smith et al.\ \cite{Smith_2018}, annotation is also frequently crowdsourced to enhance scalability and cost-effectiveness. A direct consequence of this methodology is significant acoustic variability due to differences in microphone types and recording environments. More recent approaches automate this process; for instance, MIKU-EmoBench \cite{Cheng_2025} is constructed by applying an automated pipeline to extract and label content from large-scale, user-generated video platforms. Although the acquisition is automated, this strategy retains the core benefit of crowdsourcing by capturing a wide spectrum of speech from the varied settings and diverse speaker demographics present in the original online content. While crowdsourcing and automated retrieval have expanded the scale and diversity of emotional databases, these approaches often struggle with annotation consistency, emotional ambiguity, and quality control. As a result, many large-scale corpora exhibit high variability in recording conditions and occasional inaccuracies in emotional labeling. These limitations highlight the need for frameworks that not only scale to large datasets but also maintain annotation reliability and emotion authenticity.

\subsection{Relation to Prior Work}
\label{ssec:relationpriorwork}

The effort to collect the MSP-Podcast corpus was motivated by retrieval-based strategies explored by Mariooryad \etal \cite{Mariooryad_2014_3}. The core idea was to identify emotional segments with machine learning models. We noticed this approach can scale if we design an emotion perceptual evaluation using crowdsourcing \cite{Burmania_2016_2}. Lotfian and Busso \cite{Lotfian_2019_3} formally introduced the original protocol, describing early results, showing the effectiveness of our strategy in retrieving emotional speech with the intended emotional content (e.g., finding positive speech with high valence values). Since then, we have released early versions of the corpus over the years, from version 1.0 in November 2017 to version 1.12 in June 2024. With this study, we release version 2.0 of the MSP-Podcast corpus, the final release. 

We have prepared this paper minimizing the overlap with the protocol described in Lotfian and Busso \cite{Lotfian_2019_3}. Instead, we have focused on describing the final release of the corpus and the modifications that we implemented to improve the quality of the annotations. The resulting corpus consists of 409 hours of speech, offering much broader emotional and speaker diversity than previous databases. The enhancements make the final version of the MSP-Podcast corpus a far more comprehensive and robust resource for SER research, positioning it as a superior dataset for real-world emotion recognition tasks.

\section{Protocol for the MSP-Podcast Corpus}
\label{sec:Protocol}

\begin{figure}[t]
    \centering
    \includegraphics[width=0.98\columnwidth]{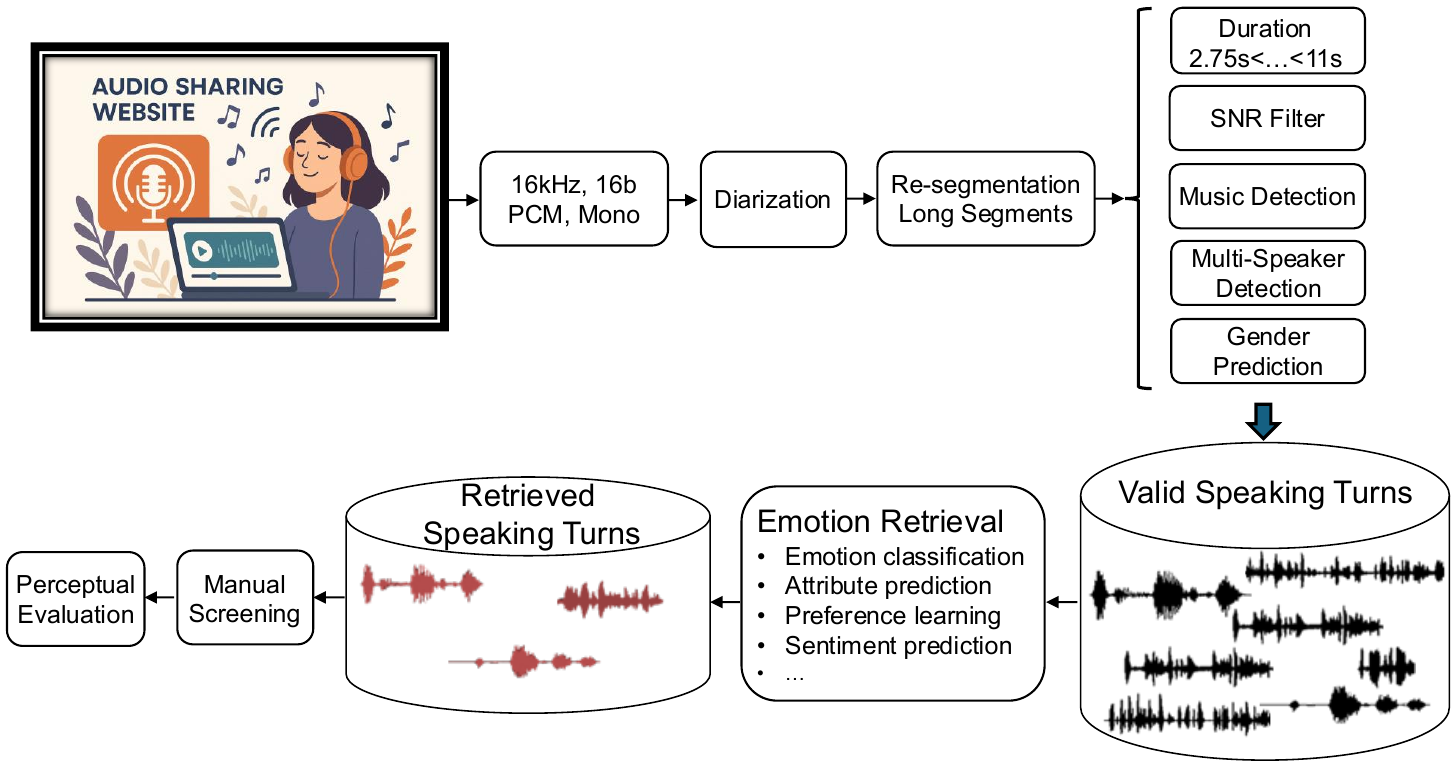}
    \caption{Protocol for the data collection of the MSP-Podcast corpus. Section \ref{ssec:selection_pods} presents the selection of podcasts. Section \ref{ssec:segmentation} discusses the data segmentation process. Section \ref{ssec:utt_selc} describes the selection of speaking turns. Section \ref{ssec:perceptualevaluation} explains the perceptual evaluation.}
    \label{fig:protocol}
\end{figure}

The protocol for data collection in the MSP-Podcast corpus is explained in Lotfian and Busso \cite{Lotfian_2019_3}. This section summarizes the protocol, with a focus on the changes implemented to enhance the quality of the data. Figure \ref{fig:protocol} shows a diagram of the data collection protocol. 

\subsection{Selection of Podcasts}
\label{ssec:selection_pods}

We source our speech data from online sources that host publicly available audio. Our goal is to have an emotionally diverse and gender-balanced corpus. We also want speaker diversity. Therefore,  we collect podcasts, talk shows, and lectures about sports, popular media, politics, personal struggles, societal issues, public health, crime, technology, and daily life. We use five criteria when searching for podcasts: (1) clean audio, no background music or speech, and not too much noise, (2) English speech, (3) emotional speech, prioritizing queries likely to convey target emotions, (4) diverse speaker demographic, and (5) appropriate license. The podcasts were identified primarily through manual searches, where researchers selected search terms that could elicit emotional topics and chose podcasts that met the aforementioned criteria. 4,743 (78.9\%) podcasts in the corpus were found in this way (manually). Eventually, we wrote a script to automatically find podcasts. A researcher can input a list of search terms, and the script will find podcasts that meet the criteria and download them. The script first downloads the metadata of some of the search results, then filters them by language (if available) and license. The script then downloads the audio of the chosen podcasts. We implement automatic steps to filter podcasts based on a music detector \cite{Lee_2017_2} and a noise detector \cite{Nicolson_2019}. Finally, a researcher briefly listens to each podcast selected by the script, verifying whether the chosen recordings meet the target criteria. 1,265 (21.1\%) podcasts in the corpus were found this way (automatically). In total, the MSP-Podcast corpus includes recordings from 6,007 unique podcasts. 

We select podcasts that are shared with licenses that allow us to distribute and modify them freely. We mainly focus on podcasts with Public Domain licenses or Creative Commons licenses with minimal restrictions (\url{https://creativecommons.org/}). Table \ref{tab:license} shows the number and percentage of podcasts in the corpus that were selected with specific licenses. Our practice was to save a screenshot of the website to document the license of the podcasts. There are 40 podcasts whose license information was not saved when initially collected, despite being selected with the target Creative Commons license. When we searched for the license information at a later date, the podcasts had been removed from the online website. Therefore, we do not have precise license information for these 40 podcasts in the corpus, which we denote as having an ``Unknown'' license in Table \ref{tab:license}.

\begin{table}[!t]
	\caption{Percentage of podcasts with a specific license in the corpus.}
	\label{tab:license}
	\centering
	\begin{tabular}{l|rrr}
		\hline
		\textbf{License} & \textbf{Perc. of Podcasts} & \textbf{\# of Podcasts}& \textbf{\# of Turns}\\
		\hline
		\emph{Public Domain} & 2.88\% & 173 & 5,872 \\
		\emph{CC-BY} & 90.86\% & 5,458 & 242,699 \\
		\emph{CC-BY-SA} & 5.59\% & 336 & 18,910 \\
        Unknown & 0.67\% & 40 & 424 \\
		\hline
        Total&--&6,007&267,905\\
  \hline
	\end{tabular}
\end{table}

After choosing and downloading the podcasts, we convert all of them to the same audio format as described in Lotfian and Busso \cite{Lotfian_2019_3}. We convert the podcasts to wave audio format with a mono channel, a sample rate of 16kHz, and 16-bit \emph{pulse code modulation} (PCM) with the Librosa toolbox \cite{McFee_2015}.

\subsection{Data Segmentation}
\label{ssec:segmentation}

The next step in the pipeline is to split the podcasts into \emph{speaking turns}. We define a speaking turn as a segment spoken by a speaker, which may comprise one or more sentences or phrases. We started the project by manually conducting this step. Researchers manually split the first 279 (4.64\%) podcasts. However, this process was very time-consuming considering the final size of the corpus. We decided to use an automated tool to split the remaining podcasts. Since podcasts can contain music or noisy segments and often feature multiple speakers, we need a tool that can segment the audio into speaking turns while also keeping speakers and noise separate. The diarization of the podcasts into sentences was mostly done using the Microsoft Azure Video Indexer \footnote{\url{https://azure.microsoft.com/en-us/products/ai-video-indexer}}. 3,667 (61.0\%) podcasts in the corpus were segmented using this tool. We eventually switched to using the Whisper model \cite{radford_2023}. 797 (13.3\%) podcasts were segmented using the pre-trained large Whisper model in the HuggingFace library \cite{Wolf_2019}. During the last part of the project, we switched to the pre-trained large-v2 Whisper model. 1,265 (21.1\%) podcasts were segmented using that model. In addition to speaker diarization, these tools provide automatic transcription of the entire podcasts.


\subsection{Automatic Filtering \& Selection of Speaking Turns}
\label{ssec:utt_selc}

After the podcasts are split into speaking turns, the next step involves employing multiple filters designed to aid our system in selecting only the highest-quality recordings to proceed with our annotation process (single speaker, no music, clean recording, with target duration, and target emotion). During this stage, we conduct several key operations: speaking turn duration estimation, resegmentation of long segments using word alignment, music detection, noise estimation, multiple speaker detection, gender prediction, automatic emotion retrieval, and final inspection by a trained human worker. This section explains each of these filters used to select the speaking turns to be included in the corpus.

\begin{figure}[t]
    \centering
    \includegraphics[width=0.9\linewidth]{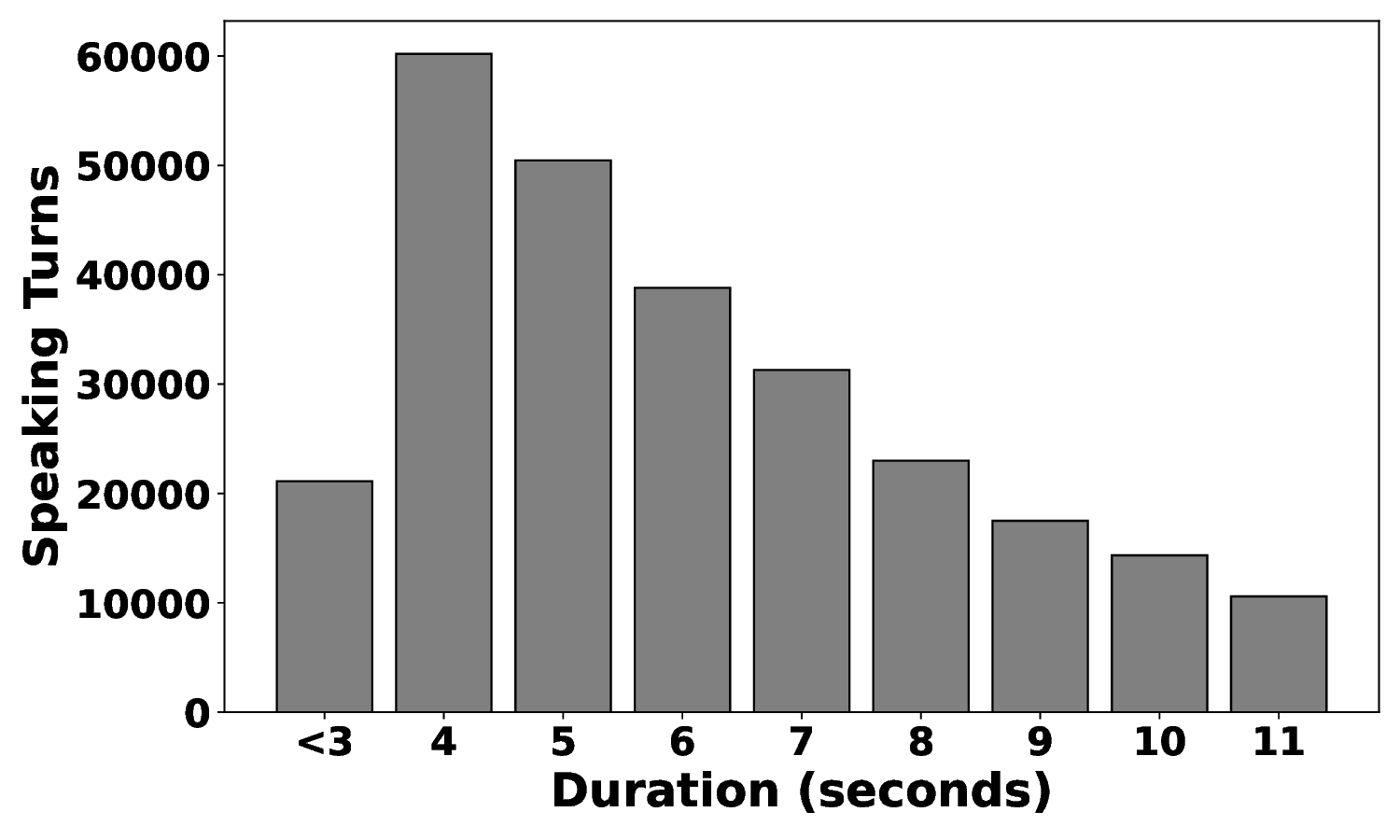}
    \caption{Histogram showing the distribution of speaking turn durations in the MSP-Podcast corpus. The x-axis shows duration in seconds.}
    \label{fig:duration_hist}
\end{figure}

The initial step involves verifying the timings and word content of the speech segments. Our goal is to have speaking turns with a duration between 2.75 and 11 seconds. The lower threshold is justified by the need to have enough context for a rater to reliably infer an emotional label during the perceptual evaluation. The higher threshold was imposed because emotions can vary during a speaking turn, so having a single label may not accurately reflect the emotional content of the speaking turn. Audios shorter than 2.75 seconds are automatically excluded, while those exceeding 11 seconds undergo a re-segmentation process. This step involves utilizing the automatic transcriptions from Section \ref{ssec:segmentation} and executing an automatic word-level alignment with the audio segments using a Python module \cite{GormanTextGrid}. This module facilitates interaction with Praat's TextGrid \cite{Boersma2001} to align transcripts with audio. We then evaluate the alignments to identify pauses in speech lasting at least 0.3 seconds, at which point we crop the audio to create smaller segments within the target range of 2.75 to 11.0 seconds. The 0.3-second threshold is applied to identify pauses indicative of a potential sentence completion by the speaker. Following this resegmentation, we aggregate all audio segments within the 2.75 to 11.0-second duration and automatically review their transcriptions to exclude any speaking turns with fewer than five words, thus eliminating segments lacking substantial spoken content. Figure \ref{fig:duration_hist} shows the distribution of the durations of the selected speaking turns included in the corpus.

The audio segments are then evaluated with music detection and noise estimation algorithms. In particular, we employ a pre-trained audio tagging model \cite{Lee_2017_2} to identify segments where music is present. Segments where music constitutes more than 50\% of the duration are filtered out. Following this step, we estimate the \emph{signal-to-noise ratio} (SNR) using the WADA-SNR algorithm \cite{Kim_2008_2}, based on \emph{waveform amplitude distribution analysis} (WADA). Audio segments with an SNR below 15dB are subsequently rejected. The remaining audio segments are further processed using the pyannote.audio speaker diarization toolkit \cite{Plaquet23,Bredin23} to ensure that each audio segment contains speech from only a single speaker. The use of this toolkit enables the automatic exclusion of samples containing multiple speakers. 

All audio segments that meet the aforementioned filters are then subjected to a series of predictive models to automatically identify speaker and recording characteristics. One of the traits is gender. Gender prediction is achieved through a pre-trained speech \emph{long short-term memory} (LSTM)-based model, capable of distinguishing between ``Female'' and ``Male'' \cite{Ertam_2019}. This process is done to gender balance the selected speaking turns.

We have millions of valid speaking turns obtained from the 6,007 podcasts that passed our criteria. Most of these segments are expected to be emotionally neutral. As explained in Lotfian and Busso \cite{Lotfian_2019_3}, we can prioritize the annotation of emotional recordings by selecting speaking turns predicted to have target emotions. Therefore, we implement an automatic emotional retrieval step. We mitigate the potential problem of biasing the selected speaking turns towards specific SER systems by employing multiple models and formulations. The SER models encompass multiple versions of emotion classification \cite{Goncalves_2022_2}, emotion attribute prediction \cite{Parthasarathy_2020, Wagner_2023}, ranking-based preference learning prediction \cite{Naini_2023}, and textual sentiment analysis \cite{barbieri_2020}. We consider open‑source implementations \cite{Wagner_2023,Parthasarathy_2020, Goncalves_2022_2, Goncalves_2024, Leem_2021, barbieri_2020, Parthasarathy_2018_3} and internally trained variants. The final retrieval system relies on over 48 criteria dictated by emotion models. It employs various pre-trained models developed from extensive emotional corpora, including CREMA-D \cite{Cao_2014_2}, MSP-IMPROV \cite{Busso_2017}, IEMOCAP \cite{Busso_2008_3}, earlier versions of MSP-Podcast \cite{Lotfian_2019_3}, and Twitter sentiment data \cite{paws_2019}. These models also utilize a comprehensive range of inputs, including \emph{low-level descriptors} (LLDs), \emph{high-level descriptors} (HLDs), raw audio for foundational \emph{self-supervised learning} (SSL) models, and textual data derived from audio transcriptions. The models were updated and retrained multiple times during the project. This emotion retrieval step is crucial for assembling an emotionally diverse and naturalistic corpus that spans a broad spectrum of emotional states. 

After running all these models on the audios, we compile a set of master lists with predictions retrieved for each task using each model, and rank these predictions from high to low accordingly for each model. We ensure that the lists are set up to dynamically change as new data is processed and entered into our master lists. Such a ranking system is instrumental in our methodology, helping us select high-emotional content and minority emotional states for annotation. Additionally, we created separate master lists for each gender. We fine-tune our selection using dynamic thresholds to maintain a balanced representation of genders and emotional states, adapting our approach as new data enters the annotation pipeline. This strategy ensures the creation of a more inclusive and precise annotated dataset, effectively minimizing bias. Updates to our master lists ensure that each sample is selected only once, avoiding redundancy in future selections. Moreover, we document the rationale behind each selection (e.g., a sample \emph{A} is chosen due to its high emotional rating by model \emph{B}), facilitating an evaluation of our models' effectiveness in identifying emotionally relevant samples for subsequent selection rounds and threshold adjustments or model removals. We weekly monitored the performance of these SER models during the project. 

Selected samples are then forwarded to a trained evaluator who conducts a thorough review, listening to each audio to confirm its suitability for annotation. This final check aims to identify any samples that, despite passing through our filters, might still present issues such as background music, low signal-to-noise ratios, unintelligible speech, foreign language usage, extremely brief sentences, profanity, multiple speakers, or excessive background noise. The evaluator's task is to identify and exclude samples based on these criteria, compiling a final list to be used for annotation. Notice that the evaluator listens only to the selected samples, instead of the millions of speaking turns in the entire pool considered for the corpus.

\subsection{Perceptual Evaluation}
\label{ssec:perceptualevaluation}

The last step in the protocol is to annotate the selected speaking turns. We annotate emotional categories (e.g., anger, happiness, etc.) and emotional attributes (valence, arousal, and dominance). Sections \ref{ssec:emo_annotation_category} and \ref{ssec:emo_annotation_attributes} describe the instrument used to annotate the corpus. The original protocol employed a slightly modified crowdsourcing strategy introduced in Burmania \etal \cite{Burmania_2016_2}. The approach tracks the quality of annotations provided by a worker in real-time during a session, stopping the session if the quality drops below a given threshold. We can measure \emph{quality} by including reference sentences that we have already annotated so that we can estimate inter-evaluator agreements. Lotfian and Busso \cite{Lotfian_2019_3} introduced specific changes to the original protocol, aiming to increase the frequency of checkpoints and incorporate primary emotional annotations and attribute-based annotations into the quality estimation. We followed this approach for the first part of the project. 

Around September 2021, we noticed important issues with our crowdsourcing platform. We noticed that \emph{human intelligent tasks} (HITs) were immediately taken when we uploaded them, suggesting the presence of bots. Several HITs returned with random annotations (e.g., all the sentences in the batch were labeled as ``happy''). Our first step was to suspend every worker found to be showing this behavior. Next, we audited and hardened the perceptual evaluation code, adding safeguards to thwart automated bot submissions and improve overall robustness. While refining our code, we developed an alternative approach to prevent delays in the annotations. We decided to hire student workers from the \emph{University of Texas at Dallas}  (UT Dallas) to annotate the corpus. Because emotion‑recognition skill varies across individuals, we created and administered a screening test to ensure we could retain only high‑performing candidates. The resulting student annotations proved consistently higher in quality than those obtained through traditional crowdsourcing. This new process enabled us to provide regular feedback to our student workers, which was not possible with crowdsourcing workers. As a result, we decided to discontinue our crowdsourcing effort and transition entirely to perceptual evaluations conducted by our student workers. Regularly, we had between 14 and 20 student workers annotating the corpus. We developed a website that connected to the server used for the perceptual evaluation, displaying the number of annotations provided by each student worker in real-time, thereby providing a powerful tool to track our progress. It was easy to identify student workers who were not actively involved in the evaluation. 

\begin{figure}[t]
    \centering
    \includegraphics[width=0.9\linewidth]{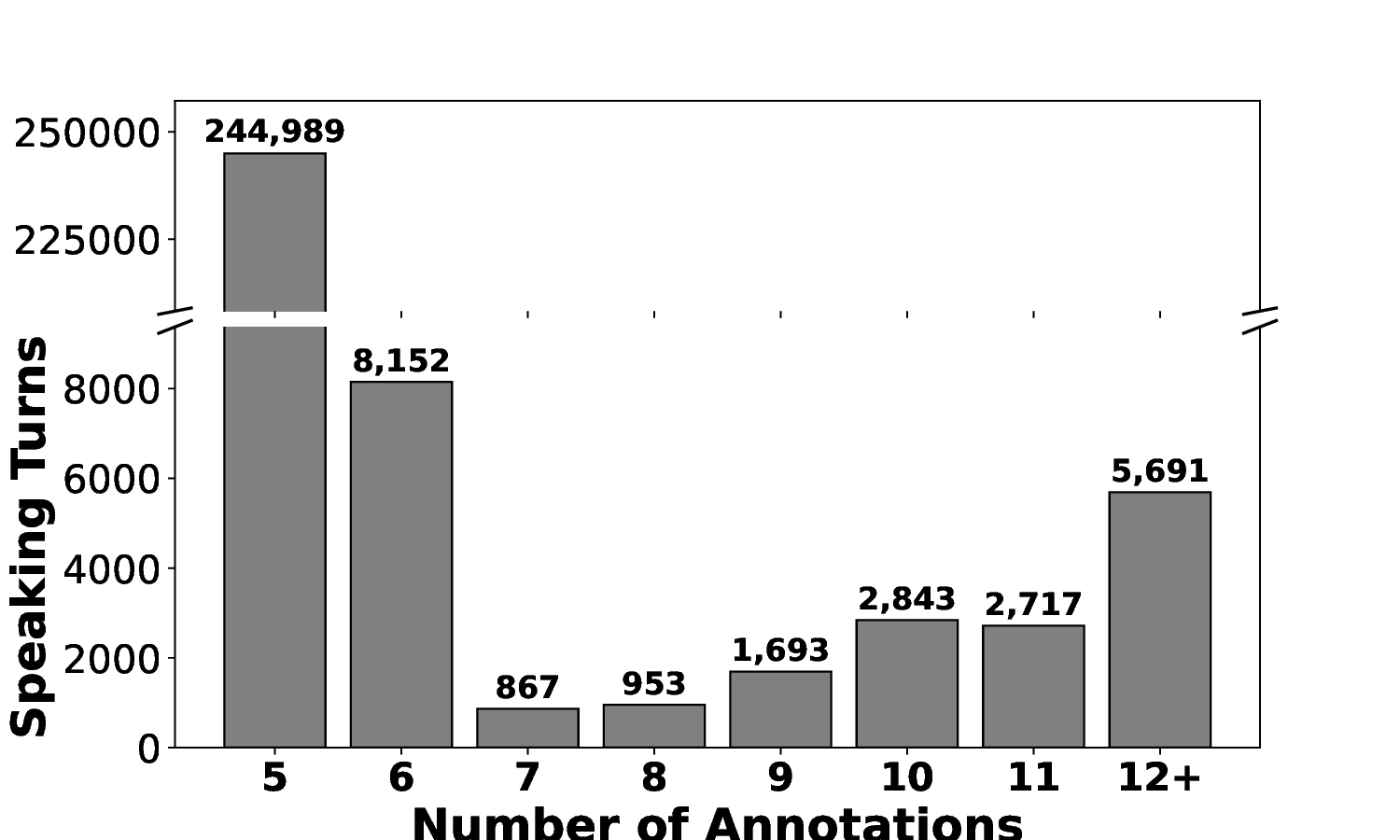}
    \caption{Histogram showing the number of files in the MSP-Podcast 2.0 corpus by the number of valid annotations. Each file has at least five annotations.}
    \label{fig:annotation_hist}
\end{figure}

We collect five or more annotations from different workers for the crowdsourcing evaluation and the perceptual evaluation conducted by our student workers. Some of the speaking turns have more than five evaluations, since they were used as reference sentences in our crowdsourcing protocol. Figure \ref{fig:annotation_hist} shows the distribution of the number of annotations per sentence in the corpus. By providing multiple annotations per speaking turn, we enable the exploration of multiple research problems related to utilizing the subjectivity of human emotional perception, such as curriculum learning training strategies \cite{Lotfian_2019_2}, exploring co-occurrence of emotion to improve the cost function \cite{Chou_2022_2}, training with soft labels \cite{Fayek_2016_2,Lotfian_2017,Sridhar_2021, Chou_2024, Chou_2025}, implementing oversampling strategies for minority classes \cite{Lotfian_2021}, and finding trends across annotations \cite{Parthasarathy_2021, Parthasarathy_2018_2}.

\begin{figure}[t]
    \centering
    \includegraphics[width=0.98\columnwidth]{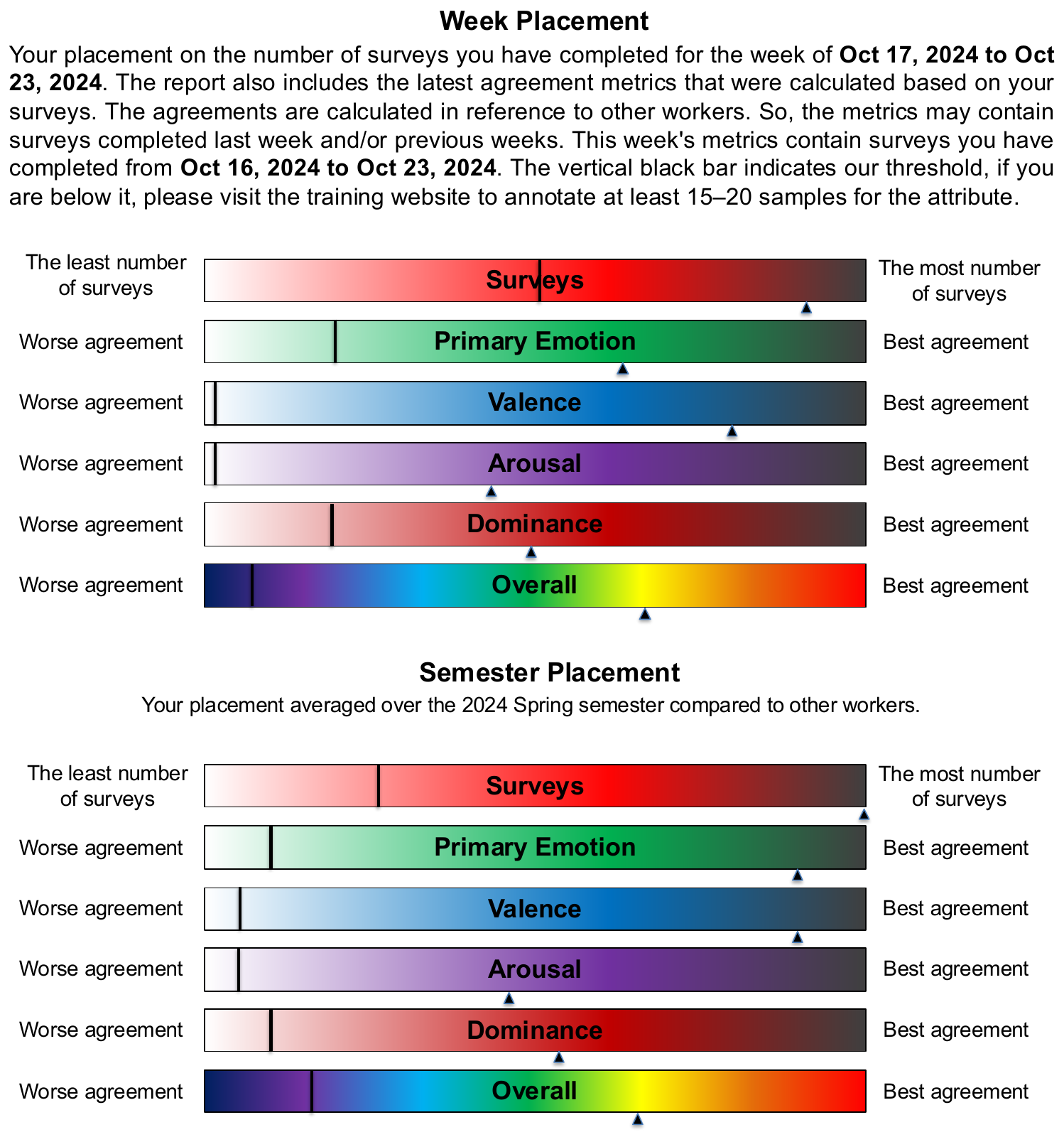}
    \caption{An example of the weekly email report sent to student workers. The email shows the number of surveys completed, their inter-agreement levels with fellow student workers, and the performance threshold (represented by the thin vertical black line) for both weekly and semester-based placements.}
    \label{fig:weekly_email}
\end{figure}

With our student workers, we did not implement the crowdsourcing strategy to track the quality in real-time. Instead, we focused on providing weekly feedback. A research assistant trained the student workers before they began annotating data, describing emotional descriptors, particularly the concepts of valence, arousal, and dominance. The student worker completed the first session with the research assistant, who answered any questions raised during the perceptual evaluation. In addition, we wanted to provide frequent feedback to the student workers, so they were aware if we were satisfied with their annotations. We implemented a weekly report that provides their relative ranking with respect to other student workers. Figure \ref{fig:weekly_email} shows an example of the document shared with our student workers. The report presents weekly-based performance (top part of the report) and semester-based reports (bottom part of the report). Instead of providing the actual values of the metrics used to estimate inter-evaluator agreements, we provide a relative ranking comparing the worker with the rest of the workers. For each indicator, we denote the performance with an arrow placed between two extremes. The closer to the right extreme, the better (see Figure \ref{fig:weekly_email}). The bars also include a black vertical line that indicates the lower threshold we tolerate. The first indicator includes the number of annotations completed by the student workers. Then, the report includes the agreement for primary emotions and attribute-based annotations (arousal, valence, and dominance). It also includes the overall score, which is the average of all the emotional descriptors. In the example in Figure \ref{fig:weekly_email}, the student worker was very good at annotating primary emotions and valence (both for the current week and the entire semester). However, the annotations for arousal and dominance were average. In all cases, the quality of the worker was above our minimum threshold. The reports were automatically generated, so this process did not require much continuous effort from our team. 

\begin{figure}[t]
	\centering
	\subfigure[Primary Categorical Emotions]
	{
	        \includegraphics[width=0.97\columnwidth]{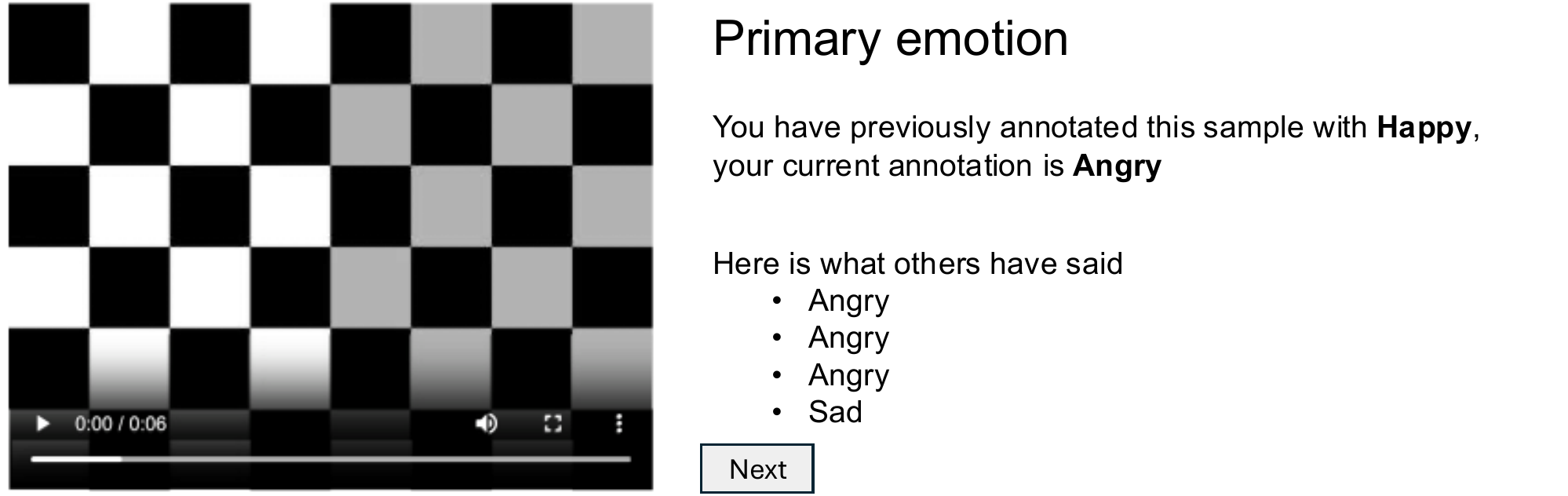}
            \label{fig:reannotations_primary}
	}
	\subfigure[Valence]
	{
            \includegraphics[width=0.97\columnwidth]{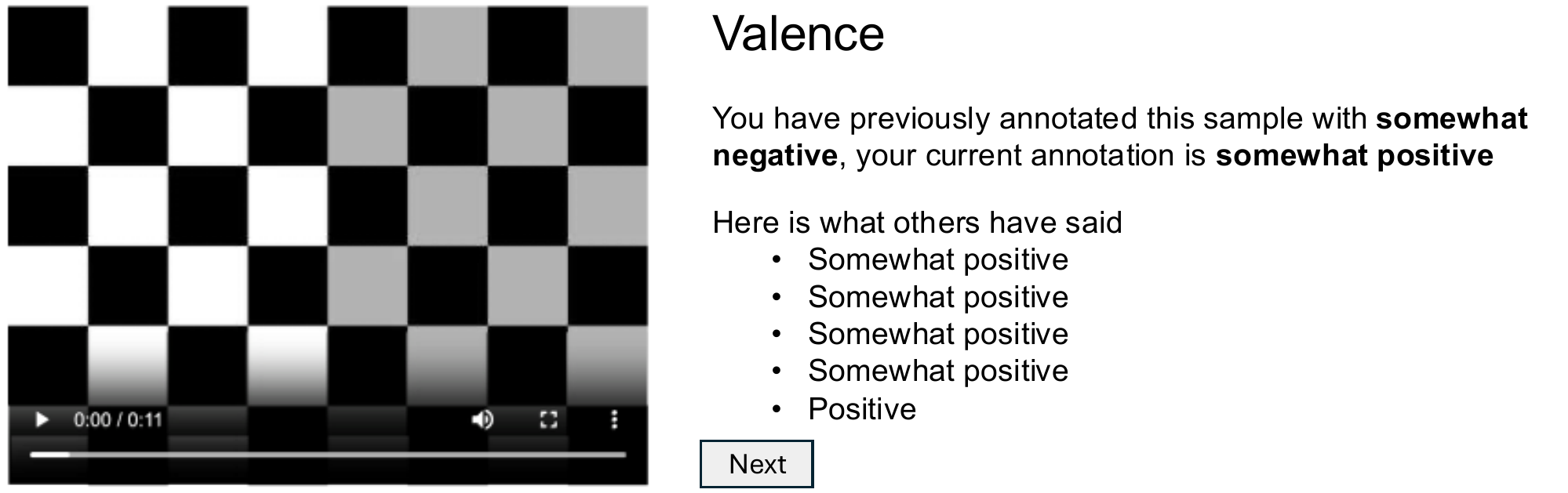}
            \label{fig:reannotations_valence}
	}
	\caption{Example of training interface for primary emotions and valence. Not shown on the image are the instructions that explain the target emotional descriptor. These screens are shown after the student worker re-annotated the carefully selected speaking turns, showing the original annotation by the target worker and the labels provided by the other student workers.}
	\label{fig:reannotations}
\end{figure}

We also implemented a targeted training to re-train our student workers with lower inter-evaluator agreements. We created a training website that focuses on a single emotional descriptor (primary emotions, valence, arousal, or dominance). Therefore, the student workers only work on the emotional descriptor that they are struggling with. For example, if a student worker has low inter-evaluator agreement on dominance, the application only includes samples to improve this emotional attribute. We automatically identify speaking turns where the annotations from the target student workers differ from consistent annotations obtained from other student workers. The application asks the student workers to re-annotate these carefully selected samples. Then, it lists their original annotations and the annotations made by the other student workers. These annotations are only revealed after the student worker re-annotates the speaking turn. Figure \ref{fig:reannotations} shows an example for primary emotions (Figure \ref{fig:reannotations_primary}) and for valence (Figure \ref{fig:reannotations_valence}). Not shown on the figures are the precise instructions given to the student workers to understand the corresponding emotional descriptors. This training was mandatory for student workers with quality below our minimum thresholds, and optional for all others who may want to practice to solidify their understanding of the emotional descriptors used in this corpus.

A later addition to the perceptual evaluation website was an optional field where a student worker could indicate that a speaking turn still had issues, despite our efforts to filter out overlapped speech, silence, noisy recordings, foreign language, or speech with background music (see bottom part of the questionnaire in Figure \ref{fig:annotations}). When a file was flagged, it was immediately separated from the perceptual evaluation until we manually checked if the speaking turn should be removed entirely from the database. This step was very important to avoid annotating data that we would later discard.

\section{Annotation of the Corpus}
\label{sec:annotation}

A key feature of the corpus is the annotations of the speaking turns. This section describes the annotations for emotions, speaker identification, human transcription, and phonetic alignment. For emotions, we utilize both categorical and dimensional attributes to describe emotions adequately.

\subsection{Annotation of Categorical Emotions}
\label{ssec:emo_annotation_category}

\begin{figure}[t]
    \centering
    \includegraphics[width=0.98\columnwidth]{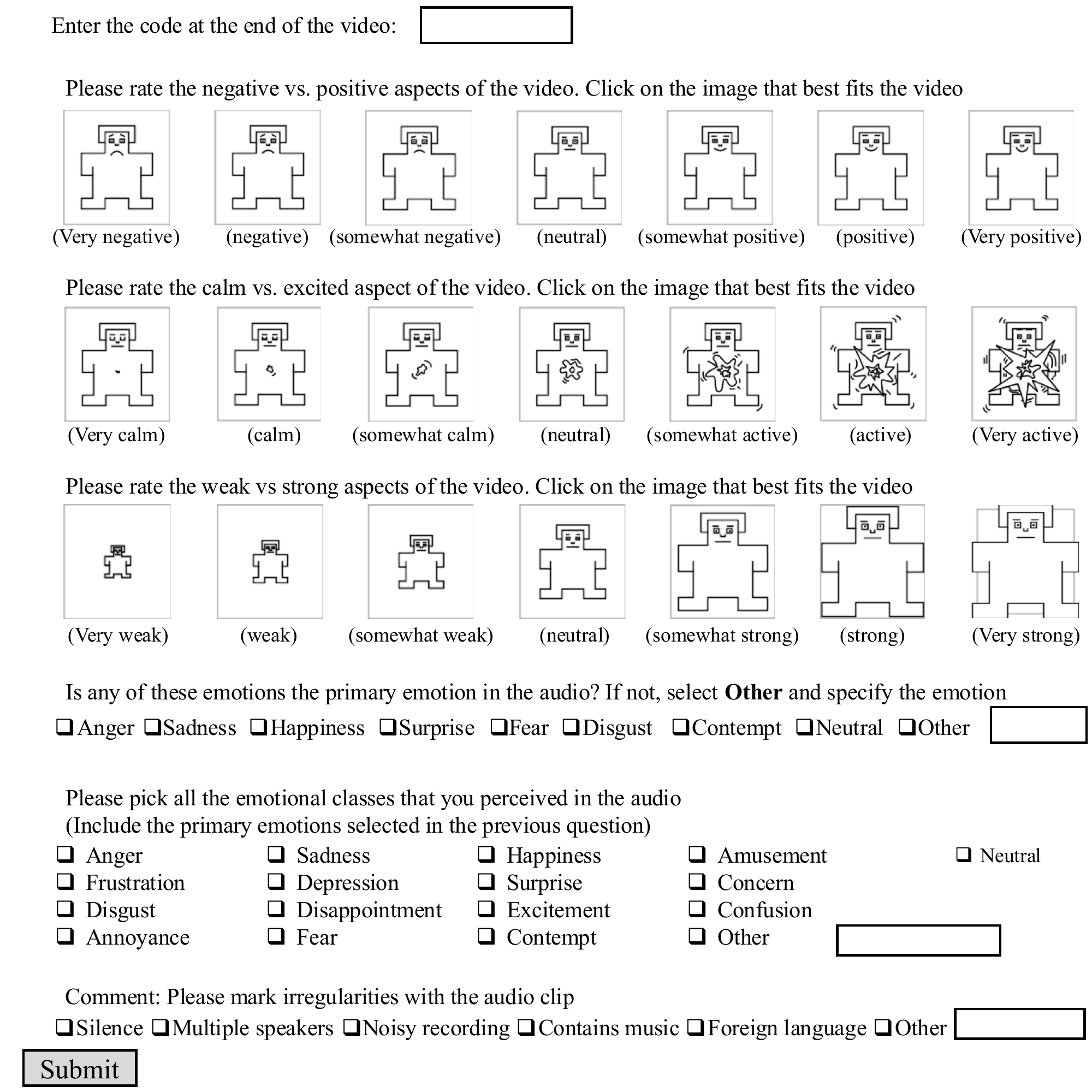}
    \caption{show the survey for annotating the MSP-Podcast audios.}
    \label{fig:annotations}
\end{figure}

The MSP-Podcast corpus offers a rich set of emotion content from natural conversational speech. Figure \ref{fig:annotations} shows the questionnaire used for the perceptual evaluation for the evaluations using crowdsourcing and student workers. The categorical annotation (bottom part in Figure \ref{fig:annotations}) was inspired by the work of Devillers \etal \cite{Devillers_2005}, which includes dominant (Major) and secondary (Minor) labels to capture mixtures of emotions. The primary emotions in the perceptual evaluation include anger, sadness, happiness, surprise, fear, disgust, contempt, and neutral speech. The workers can also select ``other'' and add their label to add flexibility and avoid the forced-choice response bias discussed by Russell \cite{Russell_1993}. The workers select only one primary emotion.

\begin{figure}[t]
	\centering
	\subfigure[Primary Emotions (consensus labels)]
	{
	    \includegraphics[width=0.98\columnwidth]{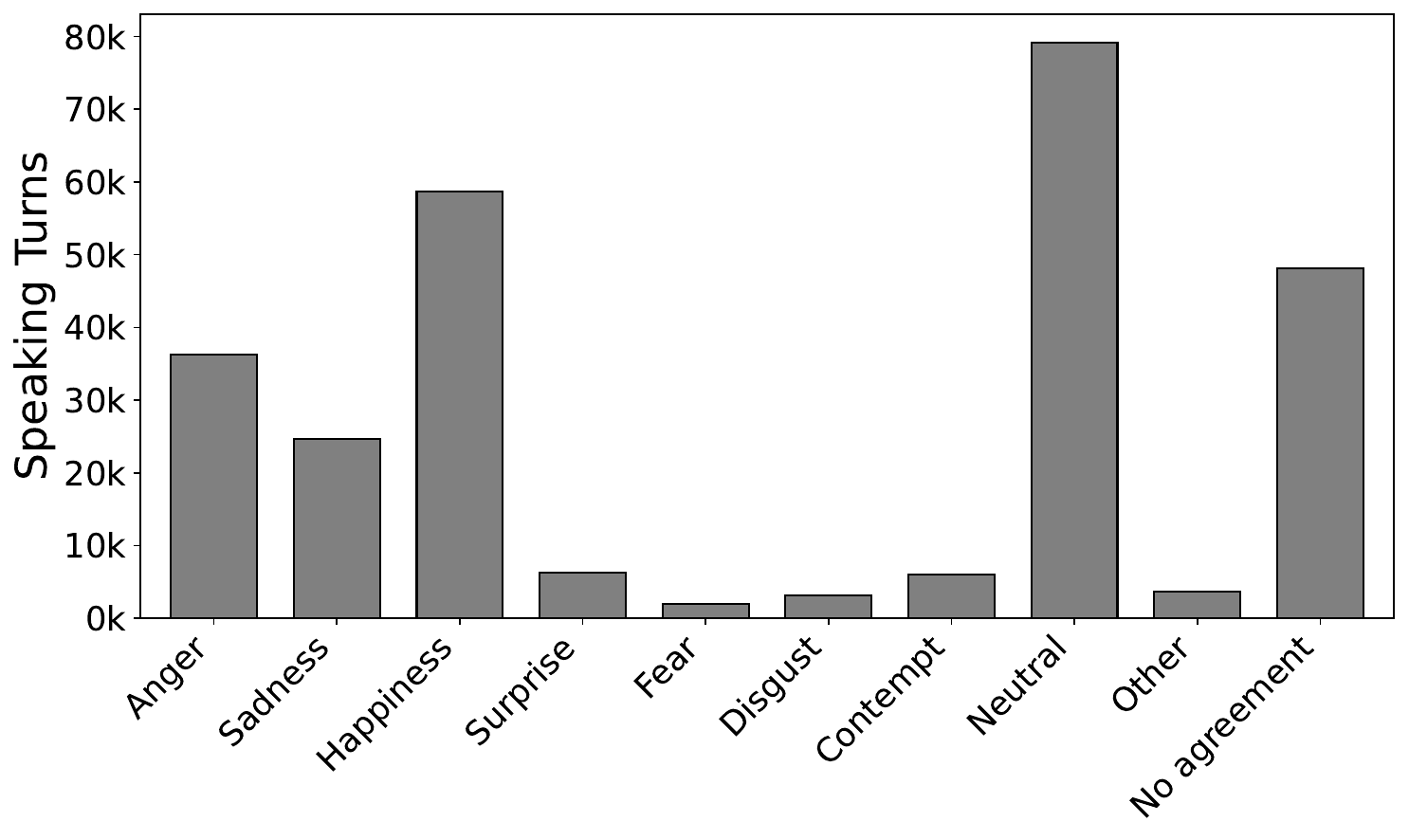}
            \label{fig:prim_emo_distribution}
	}
	\subfigure[Secondary Emotions (individual annotations)]
	{
            \includegraphics[width=0.98\columnwidth]{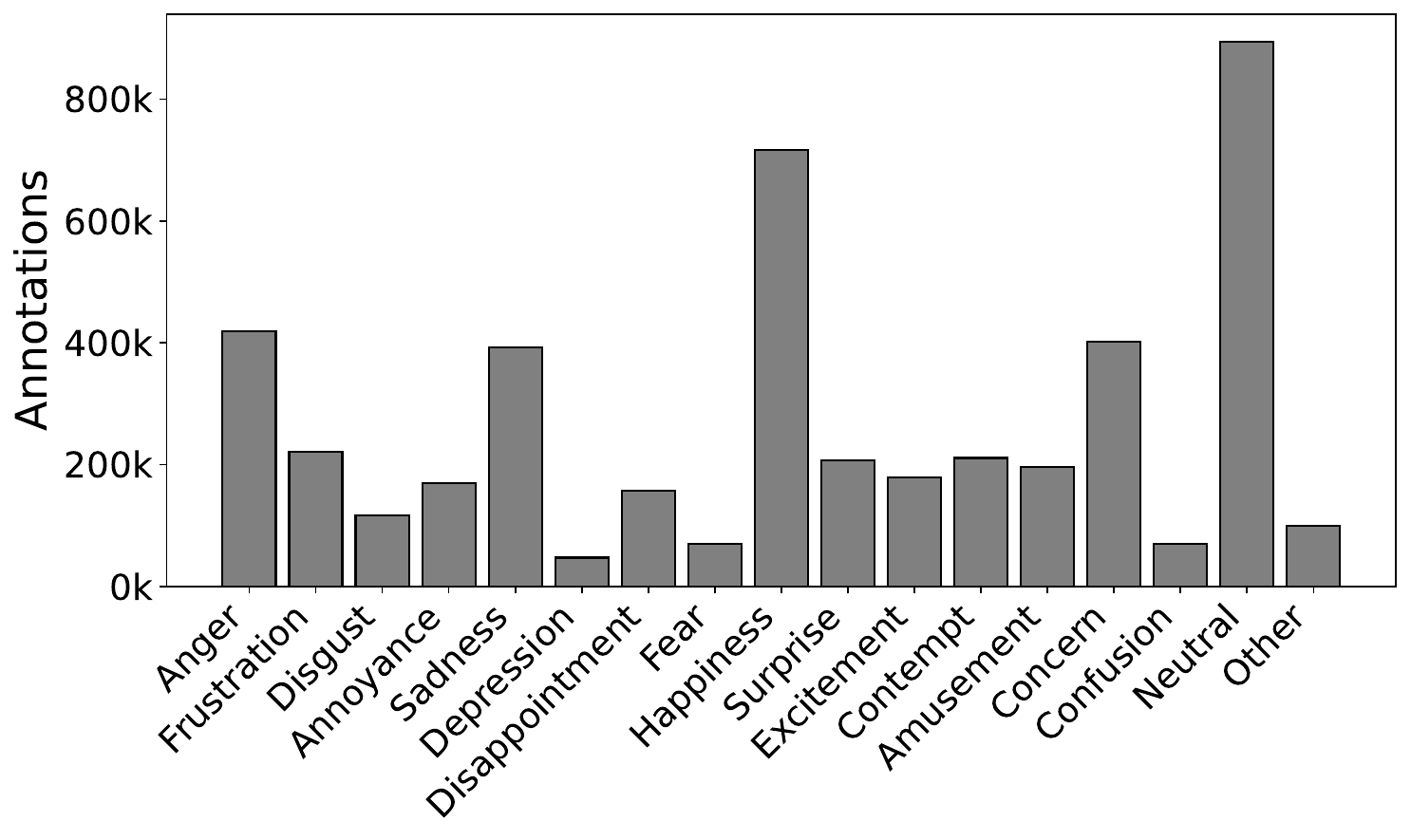}
            \label{fig:example_secondary}
	}
	\caption{Histogram of the emotional classes selected by the workers for (a) primary and (b) secondary emotions. For primary emotions, we present the consensus labels, showing the histogram of the consensus emotions assigned to the speaking turns (plurality rule). For secondary emotions, we present a histogram of the secondary emotions selected in the individual evaluations.}
	\label{fig:emo_distribution}
\end{figure}

\begin{figure}[t]
	\centering
	\subfigure[``Other'' in Primary Emotions]
	{
	    \includegraphics[width=0.95\columnwidth]{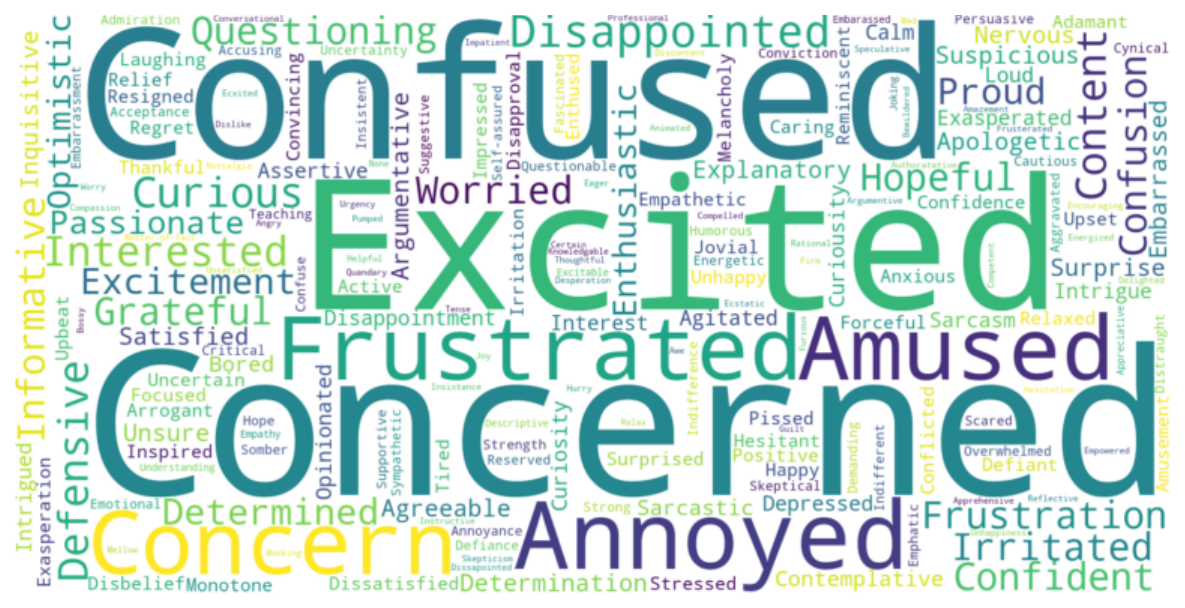}
            \label{fig:example_primary_cloud}
	}
	\subfigure[``Other'' in Secondary Emotions]
	{
            \includegraphics[width=0.95\columnwidth]{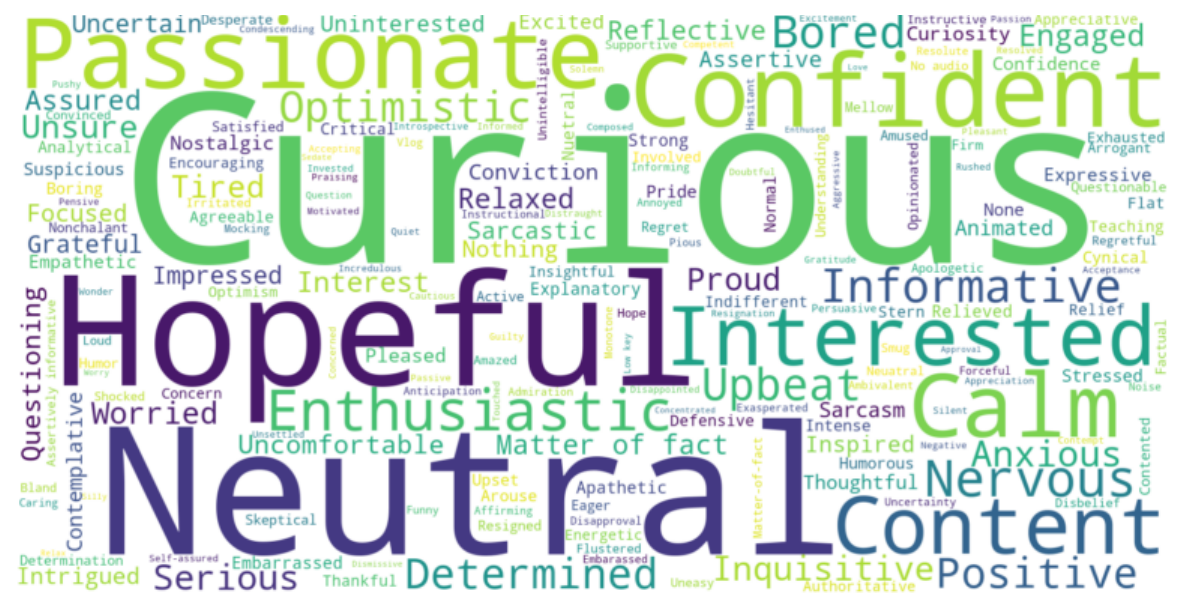}
            \label{fig:example_secondary_cloud}
	}
	\caption{Word cloud representing the manually typed emotions by the annotators when selecting the option ``other.'' (a) Other in primary emotion annotations, and (b) other in secondary emotion annotations.}
	\label{fig:example_cloud}
\end{figure}

Figure \ref{fig:prim_emo_distribution} shows the number of speaking turns assigned to each primary emotion category using the plurality rule. We include the class ``no agreement” for speaking turns that do not reach agreement under the plurality rule. The histogram reflects the frequency at which emotions appear in natural conversation, with many samples for classes such as happiness, anger, sadness, and neutral speech, and few samples for surprise, fear, disgust, and contempt. Neutral speech is the most dominant class in regular conversation. However, we only have 28\% of the speaking turns labeled as neutral, demonstrating the effectiveness of our retrieval-based strategy (Section \ref{ssec:utt_selc}). Figure \ref{fig:example_primary_cloud} shows the word cloud of the labels provided when workers selected ``other'' as the primary emotions. The figure identifies the emotions ``confused,'' ``excited,'' and ``concerned'' as the most common terms. These emotions are potential candidates for inclusion in the primary emotions for future evaluations. 

The secondary emotions extend the list of eight primary emotions by adding frustration, annoyance, depression, disappointment, excitement, amusement, concern, and confusion (16 emotions). We also include the ``other'' option, allowing them to add their own labels. The workers are asked to select all the secondary emotions that they perceived in the speaking turn. We explicitly requested that the primary class be included as one of the secondary emotions, but the workers did not always follow this instruction. Secondary emotions can play a crucial role in understanding the complex blend of emotions expressed in the speaking turns. 
Figure \ref{fig:example_secondary} shows the histogram of secondary labels selected in the individual annotations. We did not aim to obtain consensus labels like the case with primary emotions. For consistency, we added the primary emotion to the secondary emotion list when the worker did not include it. Neutral, happiness, and anger are the most commonly selected classes. If we do not include the primary emotions, the most popular selections were concern, amusement, and frustration. The classes confusion and depression were the least frequent selections.  Figure \ref{fig:example_secondary_cloud} shows the word cloud with the emotional labels provided by the workers when they selected the option ``other.'' The classes ``curious,''  ``hopeful,'' ``neutral,''  are the most frequent labels, followed by ``passionate,'' ``confident,'' ``interested,'' ``calm and ``content.'' The word cloud figures highlight the nuanced and co-occurring nature of emotions that need richer expressive descriptors to represent affective states.






\subsection{Annotation of Emotional Attributes}
\label{ssec:emo_annotation_attributes}
Emotional attributes are an alternative, powerful strategy to characterize emotions. We include the emotional attributes of valence (negative to positive), arousal (calm to active), and dominance (weak to strong). The top part of Figure \ref{fig:annotations} shows the questionnaire for these attributes. We rely on \emph{self-assessment manikin} (SAM) \cite{Bradley_1994} to visually capture the essence of each emotional attribute. We use a Likert scale from 1 to 7, with 1 indicating the lower extreme (e.g., very negative, very calm, or very weak) and 7 indicating the higher extreme (e.g., very positive, very active, or very strong). The consensus label for an attribute is the average score assigned across workers for each speaking turn.

\begin{figure}[t]
    \centering
    \subfigure[Valence]{
        \includegraphics[width=0.98\columnwidth]{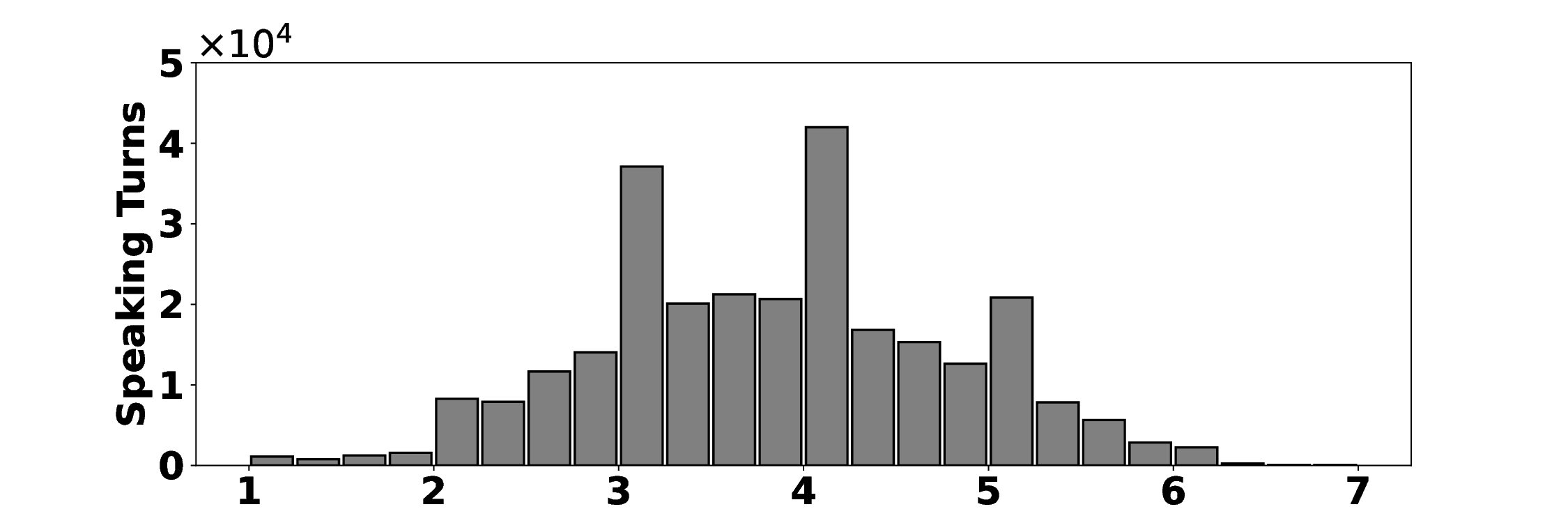}
        \label{fig:attributes_val}
    }
    \subfigure[Arousal]{
        \includegraphics[width=0.98\columnwidth]{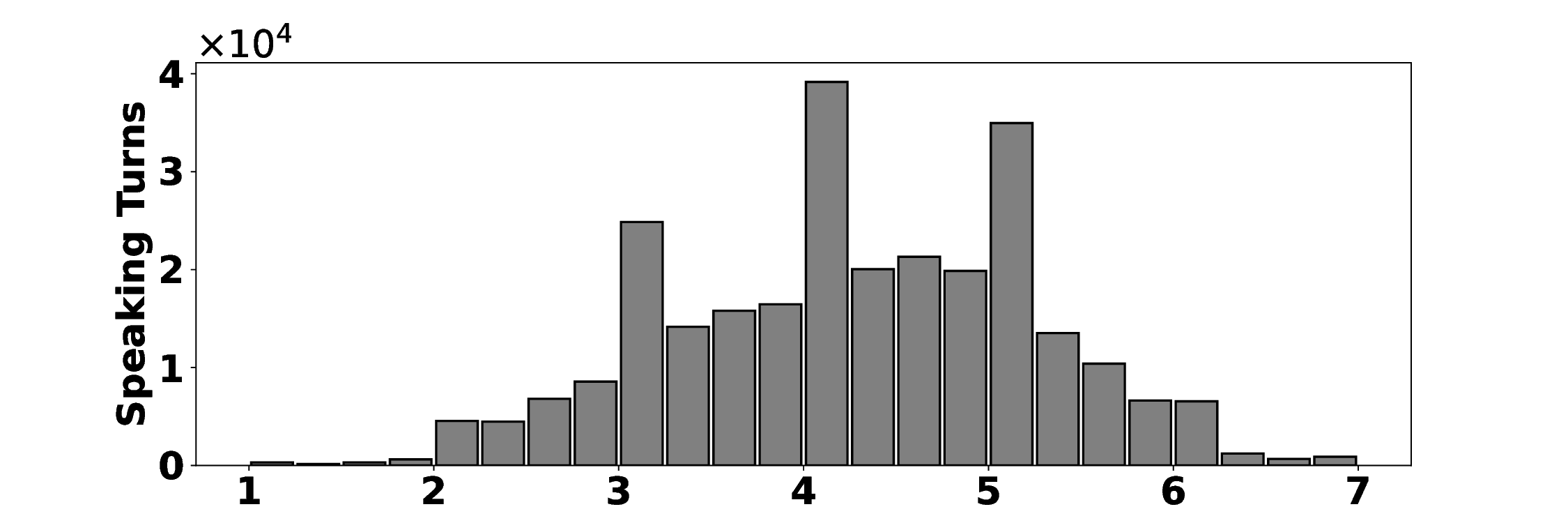}
        \label{fig:attributes_aro}
    }
    \subfigure[Dominance]{
        \includegraphics[width=0.98\columnwidth]{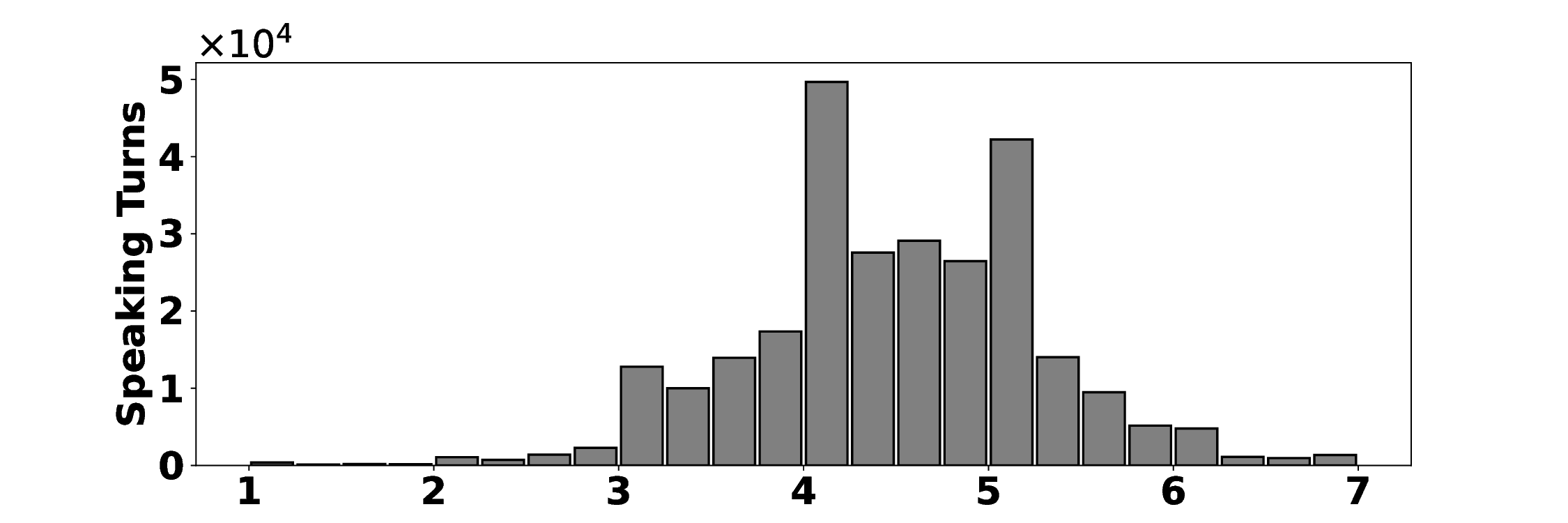}
        \label{fig:attributes_dom}
    }
    \caption{Histogram distributions of valence, arousal, and dominance attributes in the MSP-Podcast corpus.}
    \label{fig:attributes_hist}
\end{figure}

Figure \ref{fig:attributes_hist} illustrates the emotional attribute histograms of the speaking turns. Each distribution resembles a unimodal Gaussian distribution. For valence, the center of the distribution is around 4, which corresponds to the neutral range in this emotional dimension. For arousal and dominance, the distributions are slightly shifted to the right, indicating more active and dominant speech recordings. 

\begin{figure}[t]
\centering
\includegraphics[width=0.98\columnwidth]{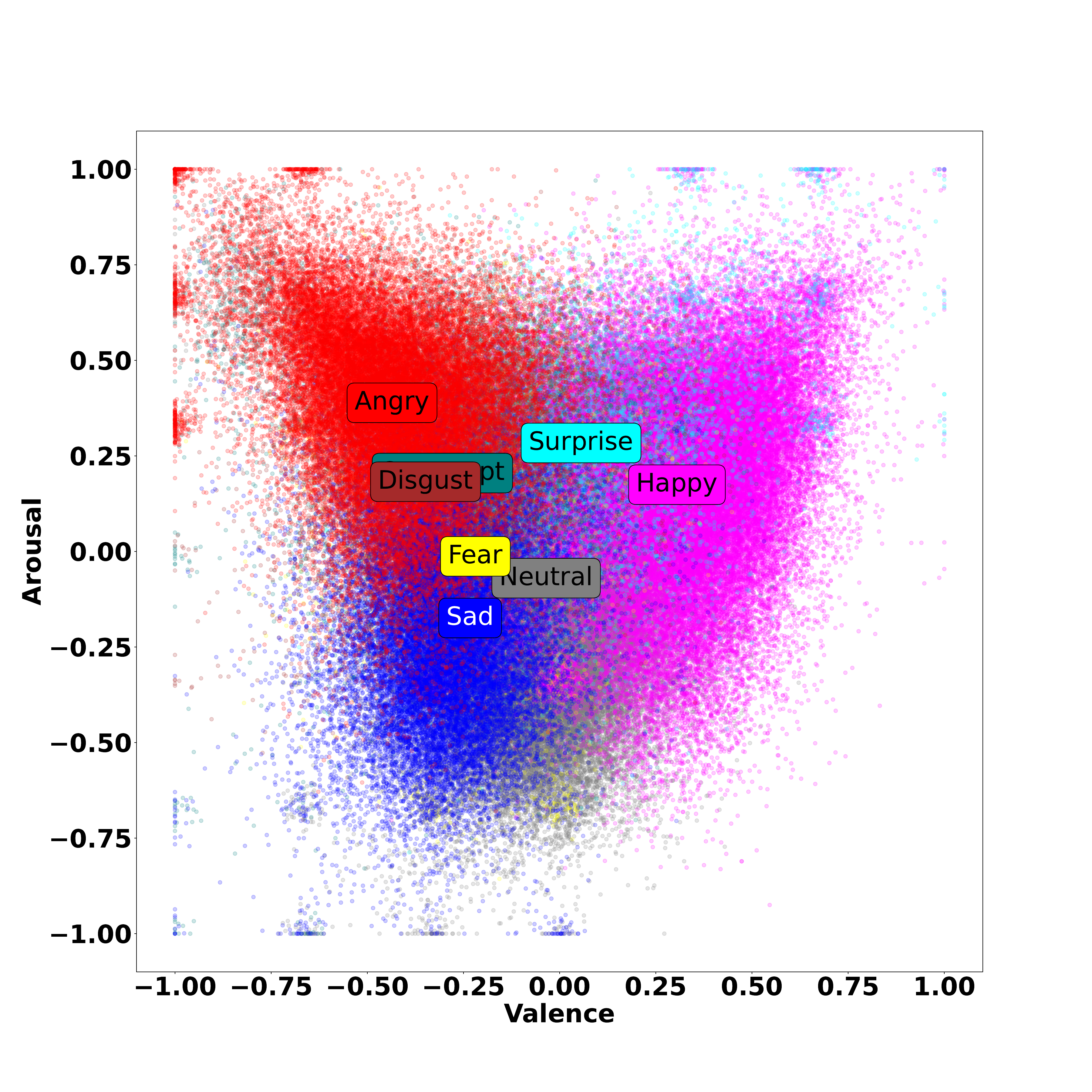}
\caption{Illustration of the emotional distribution of the MSP-Podcast corpus in the arousal and valence space, where each point is a speaking turn. The color of the points corresponds to the consensus primary class to which they are assigned. Each emotional class label is placed at the average arousal and valence values associated with that emotion. The class behind ``disgust’’ is ``contempt''.}
  \label{fig:example_valence_arousal_category}
\end{figure}

Figure \ref{fig:example_valence_arousal_category} displays each speaking turn in the arousal-valence space, with colors indicating the consensus primary emotion assigned to them. The name of each emotional class is positioned at the mean arousal and valence coordinates for that emotion. The figure shows that we have speaking turns with expressive content covering most of the arousal-valence space. The emotional classes are located in the expected quadrant of the arousal-valence space. The figure also reinforces the importance of having categorical and attribute-based annotations. We observe important intra-class variability for primary emotions, indicating that speaking turns assigned to the same class can exhibit a wide range of emotional variability (e.g., cold anger versus hot anger). By having both emotional descriptors, we can effectively capture the emotional content of the speaking turns, opening research directions that are not possible if only one of these descriptors is provided.

\subsection{Inter-evaluator Agreement}
\label{ssec:agreement}


\begin{table}[bt]
\centering
\caption{Inter-evaluator agreement in the MSP-Podcast corpus. We estimate agreement for primary emotions using Cohen's $\kappa$, and for emotional attributes using Krippendorff's $\alpha$.}
\label{tab:agreement}
\begin{tabular}{lrrrrrr}
\toprule
\textbf{Descriptor} & \textbf{All} & \textbf{Train} & \textbf{Dev.} & \textbf{Test1} & \textbf{Test2}  & \textbf{Test3} \\
\midrule
Primary [$\kappa$] & 0.411 & 0.391 & 0.410 & 0.412 & 0.294 & 0.510\\
\midrule
Valence       [$\alpha$] & 0.508 & 0.461 & 0.598 & 0.573 & 0.228 & 0.593\\
Arousal       [$\alpha$] & 0.441 & 0.412 & 0.515 & 0.471 & 0.205 & 0.610\\
Dominance     [$\alpha$] & 0.386 & 0.358 & 0.498 & 0.378 & 0.212 & 0.584\\
\bottomrule
\end{tabular}
\end{table}

Having quality emotional annotations has been a key goal of our effort. Given the struggles we experienced with crowdsourcing evaluations, we decided to estimate the inter-evaluator agreement for each worker, especially those recruited in our crowdsourcing perceptual evaluation. Based on the agreements, we removed 430 crowdsourcing workers and their 44,968 annotations. These speaking turns were reannotated with our student workers.  
After these corrections, we have 1,446,270 emotional annotations from 13,280 workers. Out of them, we have 13,205 crowdsourcing workers who completed 494,340 annotations (34.18\% of the annotations), and 75 student workers who completed 951,930  annotations (65.82\% of the annotations). The release of the corpus include the age and gender of the annotators. The inter-evaluator agreement significantly increased after re-annotating labels provided by unreliable crowdsourcing workers. The weekly feedback and the training procedure also helped improve the reliability of the labels. 

Table \ref{tab:agreement} presents the inter-evaluator agreement for the entire database and individual partitions (as described in Section \ref{ssec:Partition}). For primary emotions, the Fleiss $kappa$ statistic is 0.411 for the entire data. This agreement is high, considering the naturalness of the recordings and the inclusion of eight classes. For emotional attributes, the value for Krippendorff's $\alpha$ for valence is better than the value for arousal. Dominance is the dimension with lower agreement, although its score is above $\alpha>0.38$.

\subsection{Speaker Information}
\label{ssec:speaker_annotation}

It is essential to ensure that data splits for train, validation, and test are speaker-independent for effective SER performance that replicates the expected results on unseen data. This step requires speaker information. Knowing the identity of the speakers is also helpful to explore the role of emotions in other speech tasks such as speaker verification and identification \cite{Parthasarathy_2017_4,Pappagari_2020, Parthasarathy_2017_2,Bancroft_2019, Ulgen_2024} and speech synthesis \cite{Mahapatra_2025,Ulgen_2024_2}. Therefore, we manually annotate the speaker information of most of the corpus. 

\begin{figure}[t]
\centering
\includegraphics[width=0.98\columnwidth]{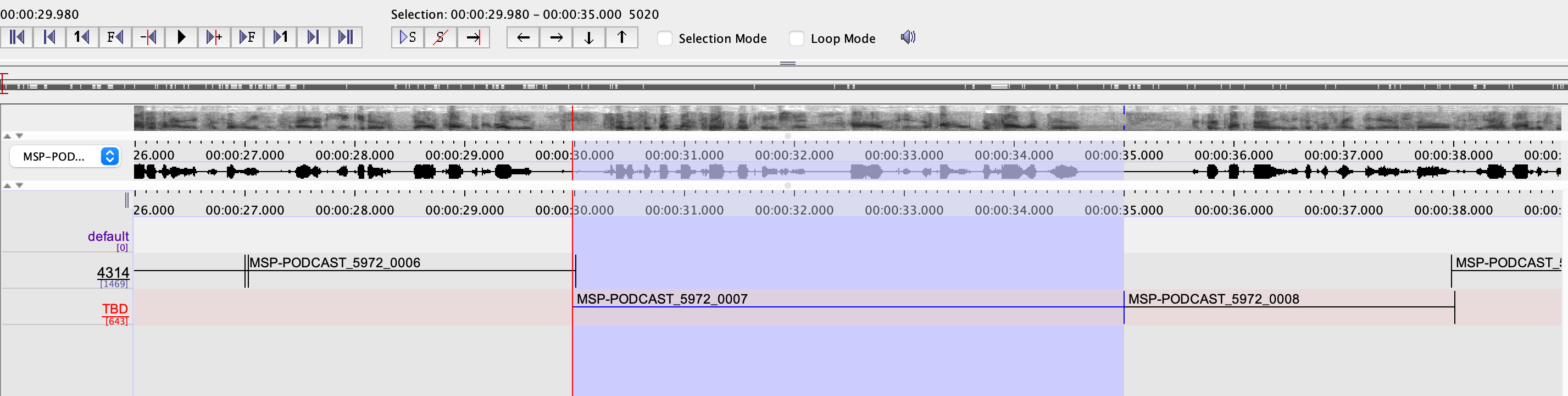}
\caption{Annotation process for speaking information using Elan. An audio track contains a previously annotated tier related to speaker 4314, providing contextual information for new annotations. A tier named `TBD' contains the speaking turns, without speaker information, to be annotated.}
  \label{fig:elan_example}
\end{figure}

As an initial step in the manual annotation process, we identify all speakers participating in a podcast session using the available information on the source webpage. Then, we listen to each speaking turn selected from that podcast and assign it to its respective speaker.
Figure \ref{fig:elan_example} shows the Elan interface used for this annotation. For each audio track, we create tiers for existing speaker annotations and a new tier for speaking turns without speaking information that we aim to annotate. As illustrated in Figure \ref{fig:elan_example}, an audio track contains two annotation tiers named `4314' and `TBD', which indicate the previously annotated speaking turns associated with speaker 4314 and the one to be annotated. We then listen to the audio around the segments, assigning speaker information to each. To maintain anonymity, each speaker is assigned a unique identification number. Some speaking turns are very hard to assign to a speaker in the conversation, even after listening to the context from nearby segments. The instruction was to mark these speakers as ``unknown,'' prioritizing precision in the annotations.    

\begin{figure}[t]
\centering
\includegraphics[width=0.98\columnwidth]{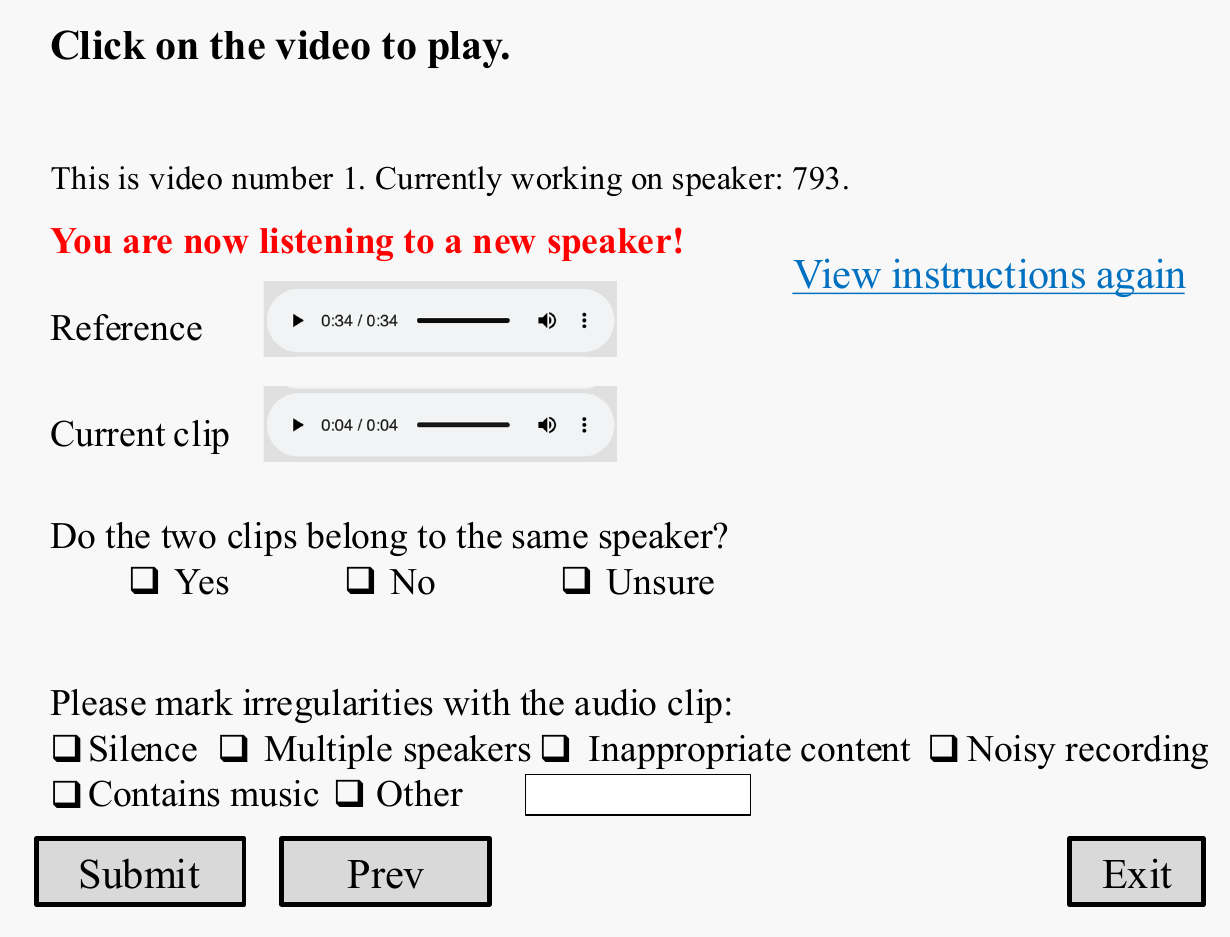}
\caption{Interface of the verification website to correct the speaker information. The annotator listens to both the reference audio and a clip that is supposed to belong to the same speaker (speaker 793 in the example). Sequentially listening all speaking turns associated with a given speaker facilitates identifying potential errors in speaker annotations.}
  \label{fig:reverify_website_example}
\end{figure}

We conducted a manual speaker verification process to correct potential mistakes made in the speaker annotations. During this process, all speaking turns associated with an individual speaker are reviewed sequentially using the user interface shown in Figure \ref{fig:reverify_website_example}. For each individual speaker, a 30-second reference audio is created by concatenating manually selected, error-free audio segments. Each speaking turn is then evaluated against this reference audio and marked to indicate whether the current clip belongs to the reference speaker. The annotators can directly compare the voice of the reference speaker with the voice of each speaking turn associated with that speaker. This method facilitates filtering outliers and inconsistencies in speaker annotations. The speaking turns flagged with wrong speaker information by this verification step are manually re-annotated to refine the speaker identities. In total, we have 3,641 unique speakers, where 2,043 are females and 1,598 are males. Table \ref{tab:NumberSpeakers} provides the number of speakers for the entire corpus and for the partitions described in Section \ref{ssec:Partition}.

\begin{table}[bt]
\centering
\caption{Speaker and gender information for the MSP-Podcast corpus. There is an overlap in the speakers included in the test sets.}
\label{tab:NumberSpeakers}
\begin{tabular}{l|rrrrrr}
\toprule
 &\multicolumn{1}{c}{\textbf{Train}} & \multicolumn{1}{c}{\textbf{Dev.}}&\multicolumn{1}{c}{\textbf{Test1}}&\multicolumn{1}{c}{\textbf{Test2}}&\multicolumn{1}{c}{\textbf{Test3}}&\multicolumn{1}{c}{\textbf{All}}\\
\midrule
Female  &1,013&298&184&53&171&1,598\\
Male    &1,207&406&281&59&257&2,043\\
Unknown &?&0&0&?&0&?\\
\hline
All     &2,220&704&465&112&428&3,641\\
\bottomrule
\end{tabular}
\end{table}

\subsection{Transcription}
\label{ssec:Transcription}


Linguistic content can provide rich information for predicting an emotion. Including text, for example, was key in recent emotion recognition challenges \cite{Naini_2025_2, Goncalves_2024}. Therefore, we provide transcription for the collected speech samples. We ask human annotators to transcribe the speaking turns in the corpus. For this purpose, we provided the collected audio files to \emph{REV.com}, which generated transcripts. Transcribers provide several indicators to describe non-verbal sounds that do not include spoken words, such as laughter or affirmative sounds. We remove indicators irrelevant to spoken information, such as \emph{(music)} or \emph{(sound)}. For consistency, we also cluster indicators that denote similar sounds, leaving eight non-verbal indicators in our transcript shown in Table \ref{tab:non_verbal_indicator}. 

\begin{table}[!t]
	\caption{Type of non-verbal indicator}
	\label{tab:non_verbal_indicator}
	\centering
	\begin{tabular}{l|rp{5cm}}
		\hline
		\textbf{Name} & \textbf{Count}&\textbf{Description}\\
		\hline
      \emph{[inaudible]} & 7,813&Unclear or unintelligible sound  \\
      \emph{[crosstalk]} &  1,970&Short overlapping speech in conversation \\
		\emph{(affirmative)} &  380&A sound indicating agreement or acknowledgment (e.g., mm-hmm, uh-huh) \\
		\emph{(negative)} & 12& A sound indicating disagreement or negation (e.g., uh-uh, mmm-mmm) \\
		\emph{(laughing)} & 78& A general laughing sound, range from soft to loud laughter \\
		\emph{(beep)} & 49& A beep sound, often indicating a censored word or alert tone \\
        \emph{(singing)} & 24& Singing voice, such as humming or melodic singing  \\
        \emph{(breathing)} & 1& A breathing sound, such as sighs or heavy breathing \\
        \emph{(cheering)} & 2& A cheering sound from crowds \\
        \hline
	\end{tabular}
\end{table}

We evaluate the quality of the annotated transcript by comparing the prediction result of robust \emph{automatic speech recognition} (ASR) systems with the collected transcript. We use OpenAI WhisperX \cite{radford_2023} and NVIDIA NeMo Canary \cite{Puvvada_2024} ASR systems for this process. We downloaded the following pre-trained checkpoints: \textit{whisper-medium.en} for OpenAI WhisperX and \textit{canary-1B} for NVIDIA NeMo Canary. These ASR systems were at the top of the rank in the Open ASR Leaderboard \cite{Srivastav_2023} (observed on Oct/23/2024). With these checkpoints, we get the ASR prediction for each speaking turn. We modified the configuration of the ASR model to make it only predicts alphabet characters without having any digits or special characters. We then compute the \emph{word error rate} (WER) between the prediction and the annotated transcript, resulting in two WERs for each of the annotated speaking turns. We ignore non-verbal indicators while computing the WER. We re-annotate transcripts for the speaking turns when both WERs are above 70\%.

\begin{figure}[tb]
    \centering
    \includegraphics[width=0.98\columnwidth]{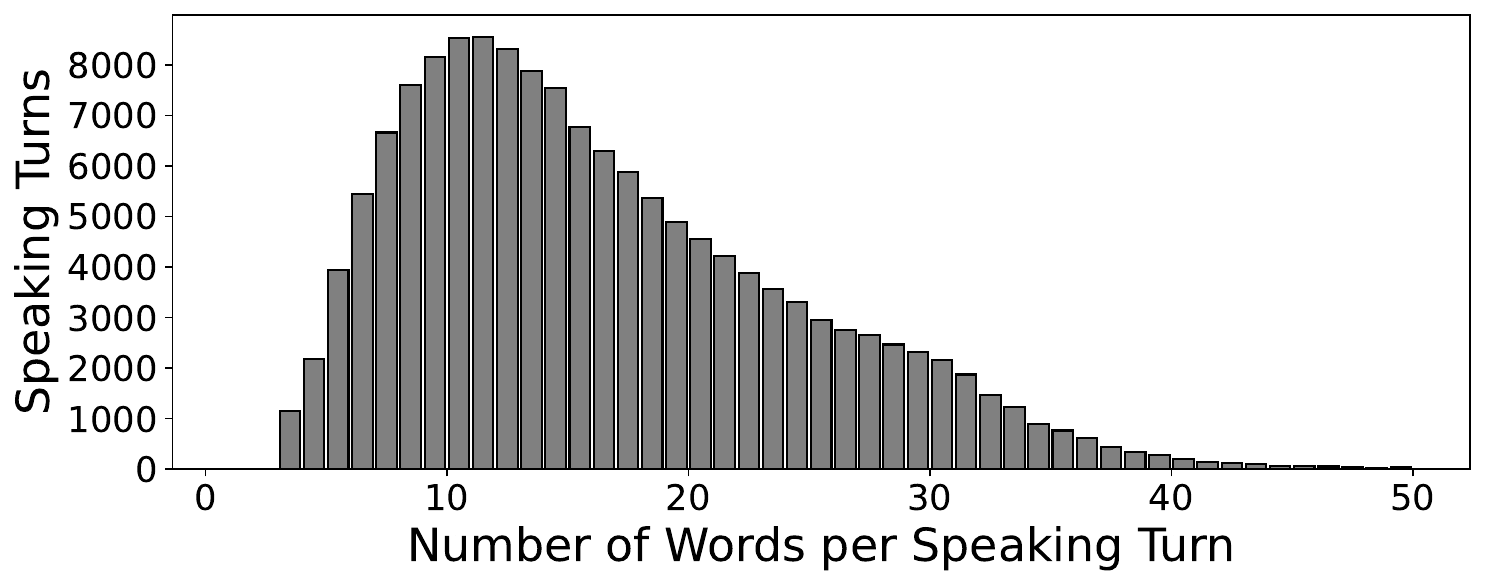}
    \caption{Histogram of number of words in the speaking turns}
    \label{fig:utterance_length_hist}
\end{figure}

The corpus contains 4.3 million tokens and 50,677 unique words, reflecting a high degree of lexical diversity. The average length of the speaking turns is 15.89 words, capturing the natural variability and spontaneity of conversational speech. Figure \ref{fig:utterance_length_hist} shows a histogram of the number of words per speaking turn, with a peak at 11 words. This distribution is consistent with conversational speech, where speakers tend to produce short but semantically rich segments.

\subsection{Phonetic Alignment}
\label{ssec:forced_alignment}
We provide time-aligned phonetic information for each speech segment in the corpus. This level of granularity enables fine-grained analysis of how phonetic structure interacts with emotions, which can support both acoustic modeling and prosody-aware emotion recognition. Importantly, these alignments facilitate cross-lingual and cross-corpus comparisons for emotion recognition, where phone-level correspondences often provide a more robust basis for knowledge transfer than lexical content alone \cite{Upadhyay_202x,Upadhyay_2023, Mote_2025}. To generate these alignments, we use the \emph{Montreal Forced Aligner} (MFA) \cite{mcauliffe2017montreal}, a widely-used tool that performs state-of-the-art alignment of phonetic units for speech given its corresponding transcript. MFA utilizes an acoustic mode implemented with \emph{Gaussian mixture models} (GMM) and \emph{hidden Markov models} (HMM). The GMM-HMM model utilizes a pronunciation dictionary to align sequences of phonemes with audio, resulting in precise timestamps for each individual phone. We used the English pretrained model and default settings provided by MFA. The resulting alignments are released in TextGrid format.

\section{Organization and Sharing of the Corpus}
\label{sec:Organization}

\subsection{Partitions}
\label{ssec:Partition}

The entire dataset is divided into multiple splits for training, development, and evaluation purposes. Table \ref{tab:splitwise_emotion} shows the distribution of primary emotions across splits. The class imbalance observed with each split is proportionally consistent with the class distribution across the whole dataset, except for \emph{test 2} and \emph{test 3}, as explained later in this section. A key distinction of our database is the addition of three test sets, which have different characteristics. The \emph{test 1} set has approximately 17.2\% of the corpus collected from 465 speakers (Table \ref{tab:splitwise_emotion}). Table \ref{tab:agreement} shows inter-evaluator agreements very similar to the values observed for the entire corpus.

\begin{table}[t]
\centering
\caption{Emotional class distribution for each partition. The MSP-Podcast corpus has 267,905 speaking turns.}
\label{tab:splitwise_emotion}
\begin{tabular}{l@{\hspace{0.25cm}}|r@{\hspace{0.25cm}}r@{\hspace{0.25cm}}r@{\hspace{0.25cm}}r@{\hspace{0.25cm}}r@{\hspace{0.25cm}}|r}
\toprule
\textbf{Emotion} & \textbf{Train} & \textbf{Dev.} & \textbf{Test1} & \textbf{Test2}  & \textbf{Test3} & \textbf{All} \\
\midrule
Anger         & 22,609 & 5,728 & 6,985 &   538 & 400 & 36,260 \\
Contempt      &  2,765 & 1,476 & 1,040 &   304 & 400 &  5,985 \\
Disgust       &  1,324 &   534 &   744 &   141 & 400 &  3,143 \\
Fear          &    794 &   285 &   348 &   116 & 400 &  1,943 \\
Happiness     & 37,048 & 7,487 & 10,948 & 2,801 & 400 & 58,684 \\
Neutral       & 51,149 & 8,318 & 12,457 & 6,793 & 400 & 79,117 \\
Sadness       & 18,256 & 2,351 &  3,041 &   581 & 400 & 24,629 \\
Surprise      &  3,220 & 1,025 &  1,206 &   394 & 400 &  6,245 \\
Other         &  1,746 &   677 &  1,019 &   277 &   0 &  3,719 \\
No agreement  & 30,279 & 6,518 &  8,506 & 2,877 &   0 & 48,180 \\
\midrule
Total         & 169,190 & 34,399 & 46,294 & 14,822 & 3,200 & 267,905 \\
\bottomrule
\end{tabular}
\end{table}

The \emph{test 2} set was collected without the retrieval-based protocol presented in Section \ref{ssec:utt_selc}. An early feedback we received was that machine learning models may bias the selection of speaking turns. We mitigate this issue by utilizing over 48 criteria based on different SER formulations, trained on different databases, features, and modalities, as explained in Section \ref{ssec:utt_selc}. In response to this problem, we also created the \emph{test 2} set. We selected 117 podcasts for this set, annotating all the speaking turns that satisfy our requirements, except the emotion retrieval step (Figure \ref{fig:protocol}). A consequence of this distinction is the higher proportion of speaking turns labeled as neutral (around 45.8\% -- Table \ref{tab:splitwise_emotion}). This test set includes recordings from 112 known speakers. An observation from this set in Table \ref{tab:agreement} is the lower inter-evaluator agreement compared to other partitions since neutral speech tends to be more uncertain \cite{Sridhar_2020}.

The \emph{test 3} set comprises 3,200 speaking turns, with a balanced representation based on primary categorical emotions (Table \ref{tab:splitwise_emotion}). These speaking turns come from 428 speakers. We are not releasing the emotional labels, transcriptions, or speaker information for this set, as it aims to provide an unbiased test set where different groups can evaluate their models and compare their results. Early versions of this test set were successfully used for SER challenges (Odyssey 2024 \cite{Goncalves_2024} and Interspeech 2025 \cite{Naini_2025_2}). We have developed a website-based interface for research groups to submit their results for classification of primary emotions and prediction of emotional attributes\footnote{\url{https://lab-msp.com/MSP-Podcast_Competition/SERB/}}. The website displays a leaderboard for each of these two SER formulations, which are automatically updated with the results of new submissions. Notice that the balance of emotional classes resulted in higher inter-evaluator agreements (Table \ref{tab:agreement}).

The development set has 12.9\% of the corpus (Table \ref{tab:splitwise_emotion}), and its purpose is to allow research teams to optimize the performance of their SER models on this set during training, including hyperparameters. This practice avoids using the test set(s) during training. The set includes recordings from 704 speakers, which are not included in either the test sets or the training set. The training set includes recordings of the remaining 2,220 speakers and the speaking turns with unknown speakers. The partitions aim to be speaker-independent, although some unknown speakers in the training set may overlap with those in the development or test partitions. The test sets should never be used for training SER models, since there is speaker overlap between test sets (e.g., data from some speakers are included in both \emph{test 1} and \emph{test 2} sets). 

\subsection{Sharing Early Versions of the Corpus}
\label{ssec:Previous}

\begin{figure}[t]
	\centering
	\subfigure[Number of Speaking Turns Over Time]
	{
	    \includegraphics[width=0.97\columnwidth, trim={0 0 0 0}, clip]{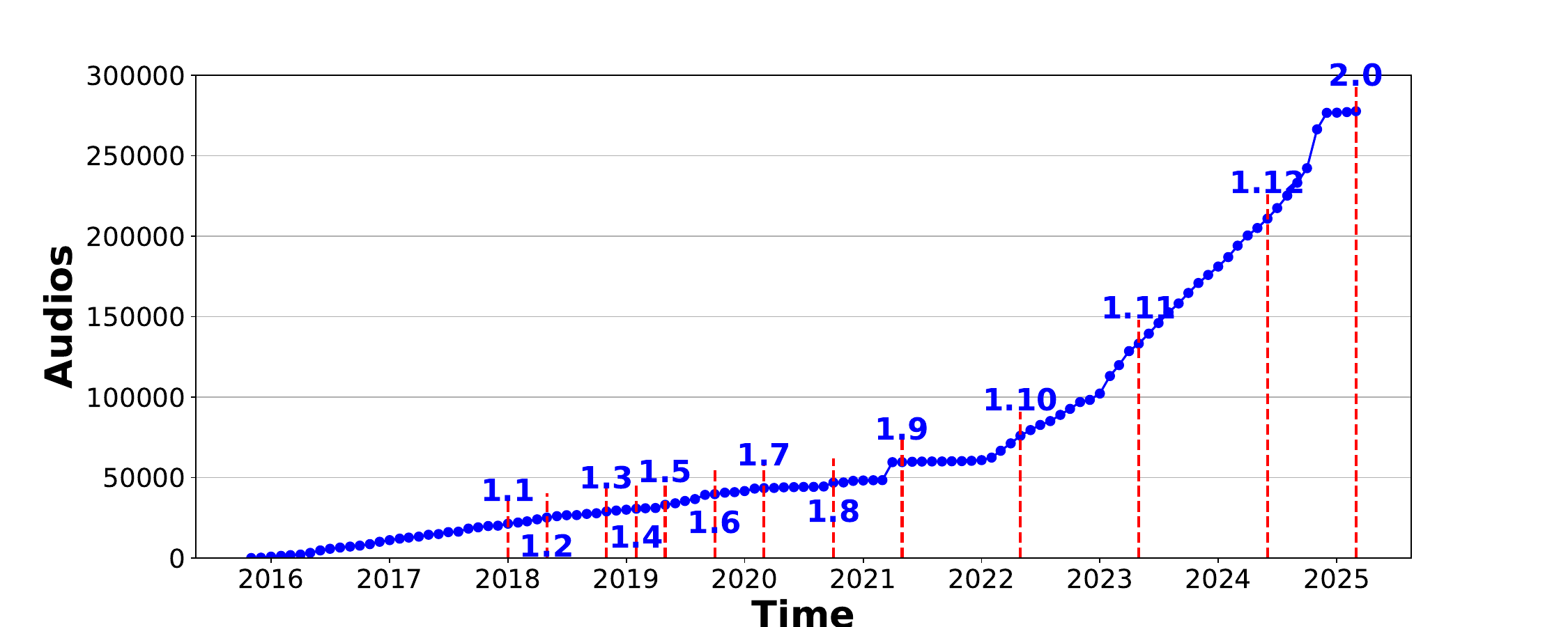}
            \label{fig:samples_timeline}
	}
	\subfigure[Number of Annotations Over Time]
	{
            \includegraphics[width=0.97\columnwidth, trim={0 0 0 0}, clip]{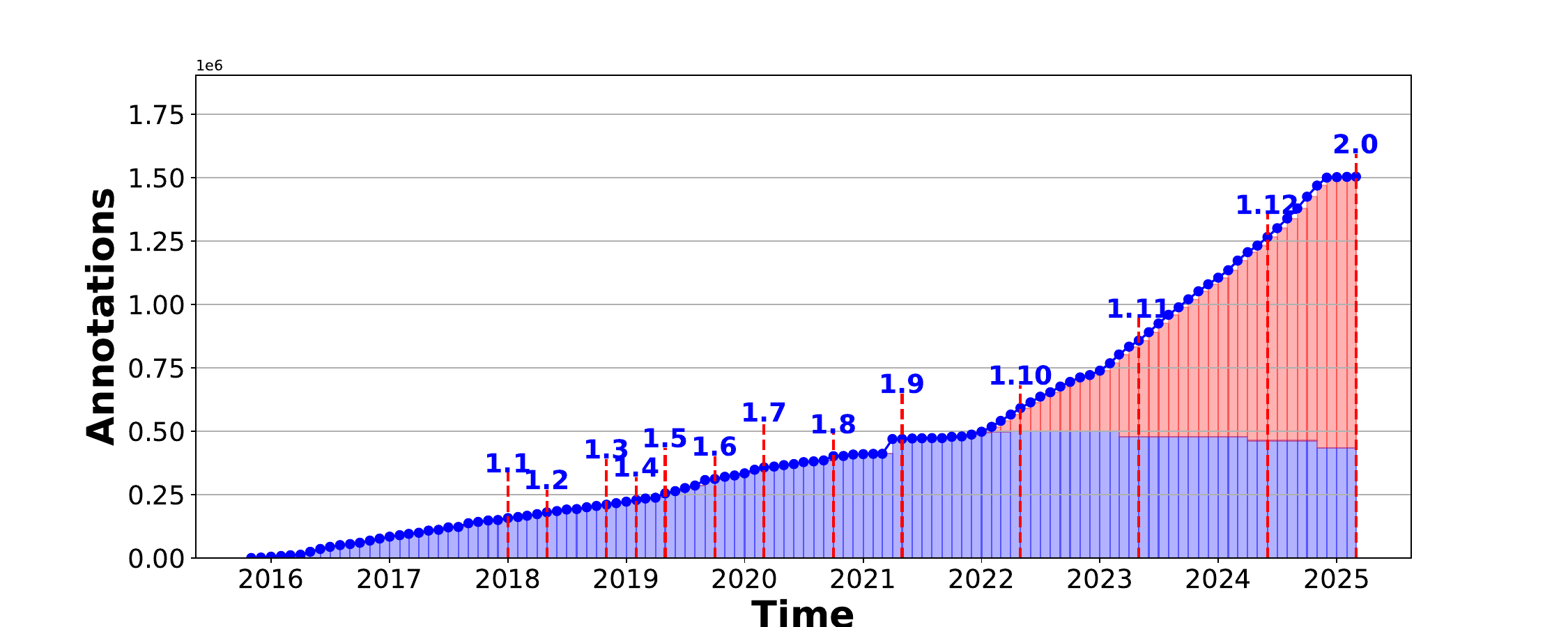}
            \label{fig:annotations_timeline}
	}
	\caption{Development of the MSP-Podcast corpus over time. The figure shows the number of (a) speaking turns and (b) annotations over time. The vertical lines indicate a released version of the corpus. For Figure \ref{fig:annotations_timeline}, the blue lines correspond to crowdsourcing evaluations and the red lines correspond to student worker evaluations.}
	\label{fig:overtime}
\end{figure}

The effort to collect the MSP-Podcast corpus started in 2015. Instead of waiting for the full corpus to be ready, we have provided partial releases so the community can benefit from this resource. Figure \ref{fig:samples_timeline} shows the number of speaking turns completed over time. The vertical lines indicate the different releases of the corpus. After transitioning to perceptual evaluations with student workers, the size of the corpus began to grow more rapidly (from 2022 to 2024). For example, in 2024 the median number of fully annotated speaking turns per week was 1,588 (up from 403 in 2020, the last year we fully relied on crowdsourcing). Figure \ref{fig:annotations_timeline} shows the number of annotations over time, indicating in blue the crowdsourcing worker annotations and in red the student worker annotations. The plot also shows an increased rate in the number of annotations from the time we fully transitioned to perceptual evaluation conducted by student workers. By the end of the project, 65.82\% of the annotations were provided by our student workers.

At the time of writing this paper, we have signed data transfer agreements with 329 academic research groups worldwide:  Africa (4), Asia (166), Australia (8), Europe (93), North America (51), and South America (7). The corpus is widely used today, playing a key role in advancing the area of speech emotion recognition.

\section{Baseline}
\label{sec:baseline}

This section presents SER results that can serve as a baseline for other researchers using this corpus. We use pre-trained SSL models built on WavLM \cite{Chen_2022}, Wav2vec 2.0 \cite{Hsu_2021_2}, or HuBERT \cite{Hsu_2021}. These models contain 24 transformer layers and are comprised of $\sim$310M parameters. We utilized the pre-trained off-the-shelf models from Hugging Face \cite{Wolf_2019}. As evidenced in previous studies \cite{Goncalves_2022_2, Wagner_2023, Wu_2024, Naini_2024, Goncalves_2024}, fine-tuning pre-trained SSL models for SER can lead to a significant performance boost. For categorical emotion recognition, we fine-tuned the models on eight emotion classes using focal loss, with a simple two-layer fully connected head. For attribute prediction, we adopted a staged fine-tuning strategy: first, adapting SSL models using \emph{concordance correlation coefficient} (CCC) loss to predict valence, arousal, and dominance, and then jointly training with categorical classification using focal loss. After the fine-tuning stage, for attribute-based predictions, we employ a single-task setup, where we train a separate regression model for each of the three emotion attributes, while keeping the SSL encoder frozen and updating only the head. We fine-tuned both models for 20 epochs, with a learning rate of 1e-5, a batch size of 32, and the Adam optimizer.

Table \ref{tab:baseline} summarizes baseline results for categorical emotion recognition and emotional attribute prediction. Overall, we observed consistent improvements across all test partitions compared to the previous MSP-Podcast v1.12 release, highlighting the benefit of expanding the training set and removing low-agreement labels. On the \emph{speech emotion recognition benchmark} (SERB) \cite{Naini_2025_2}, these refinements translated into $\sim$8\% relative gains over the earlier baselines. WavLM generally outperformed both wav2vec2 and HuBERT in both categorical and attribute tasks. The large gap between F1-macro and F1-micro scores in Test 1 reflects the severe imbalance across the eight emotion classes, where frequent categories (e.g., neutral, happiness) dominate the micro-average. These results provide a stronger and more reliable baseline for future work in categorical and dimensional SER.

\begin{table}[t]
\centering
\caption{Baseline performance on categorical emotion recognition and emotional attributes recognition.}
\label{tab:baseline}
\vspace{2mm} 
\begin{tabular}{l@{\hspace{0.25cm}}|c@{\hspace{0.25cm}}c|c@{\hspace{0.25cm}}c|c@{\hspace{0.25cm}}c}
\multicolumn{7}{c}{\textbf{Categorical Emotions}} \\
\hline
&\multicolumn{2}{c|}{\textbf{Test 1}}& \multicolumn{2}{c|}{\textbf{Test 2}}&\multicolumn{2}{c}{\textbf{Test 3}}\\
\hline
Model & F1-Ma & F1-Mi& F1-Ma & F1-Mi& F1-Ma & F1-Mi \\
\hline
WavLM & 0.297 & 0.394 & 0.206 & 0.280 & 0.356 & 0.373\\
Wav2vec 2.0 & 0.238 & 0.325 & 0.156 & 0.166 & 0.289 & 0.316\\
HuBERT & 0.285 & 0.390 & 0.192 & 0.264& 0.344 & 0.361\\
\hline
\end{tabular}

\vspace{5mm} 

\begin{tabular}{cl|ccc}
\multicolumn{5}{c}{\textbf{Emotional Attributes}}\\ 
\hline
&Model & Valence & Arousal & Dominance\\
\hline
\multirow{3}{*}{\rotatebox{90}{\textbf{Test 1}}}
&WavLM & 0.722 & 0.724 & 0.645 \\
&Wav2vec 2.0 & 0.692 & 0.718 & 0.639  \\
&HuBERT & 0.720 & 0.708 & 0.648 \\
\hline
\multirow{3}{*}{\rotatebox{90}{\textbf{Test 2}}}
&WavLM & 0.549 & 0.547 & 0.467 \\
&Wav2vec 2.0 & 0.479 & 0.553 & 0.467 \\
&HuBERT & 0.541 & 0.533 & 0.465 \\
\hline
\multirow{3}{*}{\rotatebox{90}{\textbf{Test 3}}}
&WavLM & 0.632 & 0.632 & 0.479 \\
&Wav2vec 2.0 & 0.625 & 0.634 & 0.476 \\
&HuBERT & 0.641 & 0.630 & 0.489 \\
\hline
\end{tabular}
\end{table}

\section{Discussion}
\label{sec:discussion}

The MSP-Podcast corpus opens new research possibilities due to its unique features, including its diversity in speakers, emotions, and environments. Wagner \etal \cite{Wagner_2023} and Naini \etal \cite{Naini_2024} demonstrated that finetuning SSL models such as WavLM with emotional data is beneficial for SER tasks. This corpus is sufficiently large to support effective finetuning, providing a stronger starting point for models tailored to a specific domain where less annotated data may be available. This database unlocks a range of novel opportunities. We focus here on highlighting a few notable ones.

\subsection{Perception of Emotions}
\label{ssec:perceptionfw}

With 1,446,224 annotations from 13,278 workers, this corpus is well-suited for studying human emotion perception. We are releasing all individual annotations, along with the timestamps indicating when each annotation was completed. This information enables research that incorporates contextual factors into emotion perception. For instance, it allows investigation of the priming effect -- how previously annotated sentences influence the perception of subsequent speaking turns \cite{Martinez-Lucas_2023,Martinez-Lucas_202x}. The sequential order of the annotations can also support preference learning strategies, where direct comparisons are used to establish relative labels (e.g., one speaking turn is more positive than another) \cite{Naini_2023_2}. 

A related resource is the MSP-Conversation corpus \cite{Martinez-Lucas_2020}, which includes time-continuous annotations of 10–20 minute segments from the same podcasts used in the MSP-Podcast corpus. These annotations provide continuous traces of perceived changes in valence, arousal, and dominance over time. There are 12,561 segments in the MSP-Podcast that overlap with the recordings in the MSP-Conversation corpus. This overlapping set offers an opportunity to study the relationship between continuous-time annotations (MSP-Conversation) and sentence-level annotations (MSP-Podcast) \cite{Martinez-Lucas_2024_2}.

\subsection{Robustness to Environments}
\label{ssec:noisefw}

The variety of podcasts used in this corpus provides a perfect resource for evaluating speech models that are robust to multiple environments. We highlight two prominent efforts in this area. Leem \etal \cite{Leem_2021} recorded an early version of the MSP-Podcast corpus by playing the speaking turns and radio noise in a single-walled sound booth (release 1.8). The microphone and the speaker were strategically placed at different locations to achieve target SNRs. This noisy version of the corpus has been extremely useful to explore robust SER models \cite{Leem_2022}. The second effort is the work of Grageda \etal \cite{Grageda_2023,Grageda_2025}, which recorded a noisy version of the MSP-Podcast corpus in the context of \emph{human robot interaction} (HRI) (\emph{test1} of release 1.9). The microphone was mounted on a robot, which moved, changing the relative distance between the noise source, the speech source, and the microphone. This effort has led to improvements in distant SER models \cite{Garcia_2024}. 

\subsection{Emotions and Other Speech Tasks}
\label{ssec:}

\begin{figure}[t]
    \centering
    \includegraphics[width=0.95\columnwidth]{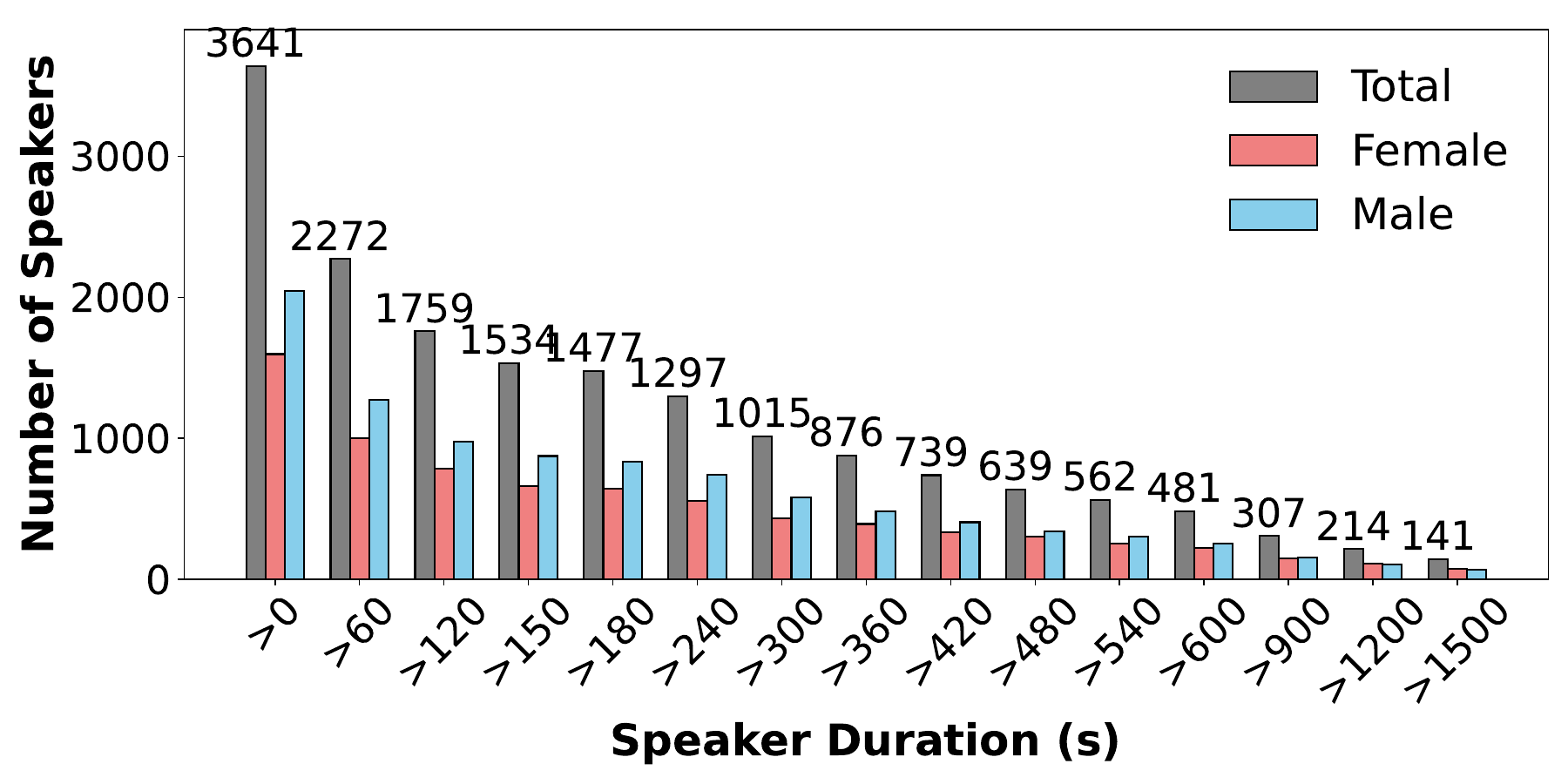}
    \caption{Cumulative distribution of speakers with increasing recording duration. The bars show the number of male, female, and total speakers who have more than a given duration of data in seconds.}
    \label{fig:speaker_duration_distribution}
\end{figure}

The size of the corpus and the speaker information make this resource ideal for exploring how emotion affects other speech tasks, such as speaker verification and speaker recognition tasks \cite{Parthasarathy_2017_4,Pappagari_2020, Parthasarathy_2017_2, Bancroft_2019, Ulgen_2024}. To support these tasks, we made a key decision to collect multiple podcast episodes from the same speakers whenever possible. Speaker verification evaluations are often conducted across sessions collected on different days under different conditions. Different episodes are often collected on different days, which approaches this evaluation setting where several speakers appear in multiple podcasts. Likewise, many applications and experimental settings require sufficient recordings from individual speakers, which we ensured by including multiple episodes per speaker. For example, speaker verification tasks require an enrollment set to build the models. Also, \emph{text-to-speech} (TTS) requires enough data to build a speaker model. Figure \ref{fig:speaker_duration_distribution} shows an accumulative plot with the number of speakers having a given amount of data. For example, there are 1,015 speakers with 300 seconds (5 minutes), and 141 speakers with 1,500 seconds (25 minutes) of data. These features make this corpus ideal for \emph{voice conversion} (VC) and TTS tasks \cite{Mahapatra_2025,Ulgen_2024_2}.

\subsection{Rich Emotional Descriptors}
\label{ssec:secondary}

Most emotional corpora provide either categorical or attribute-based annotations. In contrast, the MSP-Podcast offers both, along with secondary emotion labels, where annotators select all emotions they perceive in a recording. We have shown the value of secondary emotions by using them as auxiliary tasks in classification problems \cite{Lotfian_2018}, and in retrieval tasks aimed at finding recordings with emotions similar to a reference (anchor) sample \cite{Harvill_2023,Harvill_2019}. As described in Section \ref{ssec:emo_annotation_category}, the annotation protocol allows evaluators to provide their own labels for both primary and secondary emotions when none of the predefined options are appropriate. This information is also valuable, as demonstrated by Chou \etal \cite{Chou_2022}, who transformed the free-text labels into polarity vectors (negative, positive, ambiguous) using LIWC \cite{Pennebaker_2015}. These examples showcase the potential of the rich emotional descriptors provided in the corpus. 

\subsection{Support for Other Data Collections}
\label{ssec:otherdatafw}

The focus of this project is on speech recordings in English. There is a need to collect similar databases in other languages. We created the \emph{affective naturalistic database consortium} (AndC)\footnote{\url{http://andc.ai/}}. This initiative aims to provide all the tools used to collect the MSP-Podcast corpus to other researchers, enabling them to create new databases and expand the infrastructure for affective computing. We have partnered with collaborators from the National Tsing Hua University in Taiwan to test this initiative. They followed the code and protocol used for our corpus. The result of this effort is the BIIC-Podcast corpus \cite{Upadhyay_2023_2}, with recordings in Taiwanese Mandarin. Another example is the collection of the \emph{White House tapes speech emotion recognition} (WHiSER) corpus \cite{Naini_2024_2}. Using a variation of the proposed protocol, we annotated the emotions of ambient recordings from the Oval Office during the presidency of Richard Nixon. This set provides a perfect test set for SER models in challenging recording conditions (distant speech, low-quality microphones, noisy environment). We expect that this consortium will encourage the creation of new resources.

Another collaboration that started from this effort is the NaturalVoices corpus \cite{Salman_2024,Du_2025}. This database uses the 6,007 recordings used in the MSP-Podcast corpus (5,046 hours). While MSP-Podcast was originally developed for SER, NaturalVoices is tailored for speech generation tasks, particularly \emph{voice conversion} (VC) \cite{Salman_2024} and \emph{emotional voice conversion} (EVC) \cite{Du_2025}. Its annotations and data processing pipeline are specifically designed to support these tasks, although the corpus is also suitable for other speech synthesis applications such as \emph{text-to-speech} (TTS). The original podcast recordings are freely available \footnote{\url{https://github.com/3loi/NaturalVoices}}. The MSP-Conversation corpus \cite{Martinez-Lucas_2020} also beneficed from the collection of the MSP-Podcast corpus.

\section{Conclusions}
\label{sec:conclusion}
This paper presented the results of a 10-year effort to develop the MSP-Podcast corpus -- a large, naturalistic emotional speech database containing diverse recordings from multiple speakers across various environments. The database reflects the emotions observed in daily human interactions. The corpus includes a rich set of emotional descriptors, enabling new research in emotion analysis, recognition, and synthesis. To ensure high-quality annotations, we implemented several strategies, including a screening test for student workers prior to hiring, weekly feedback, and targeted training to improve consistency in labeling. In addition to releasing the final version of the corpus, we also provide the code used in the protocol (Section \ref{ssec:otherdatafw}), with the intention of supporting replication efforts that will expand affective computing resources in other languages.


%

\ifCLASSOPTIONcompsoc
  \section*{Acknowledgments}
\else
  \section*{Acknowledgment}
\fi
We are grateful to the more than 13,720 individuals who contributed to this effort. We use AI systems for editing and grammar enhancement.

\ifCLASSOPTIONcaptionsoff
  \newpage
\fi



%

\bibliographystyle{IEEEtran}
\bibliography{reference}

%

\vspace{-1.3cm}
\begin{IEEEbiography}[{\includegraphics[width=1in,height=1.25in,clip,keepaspectratio]{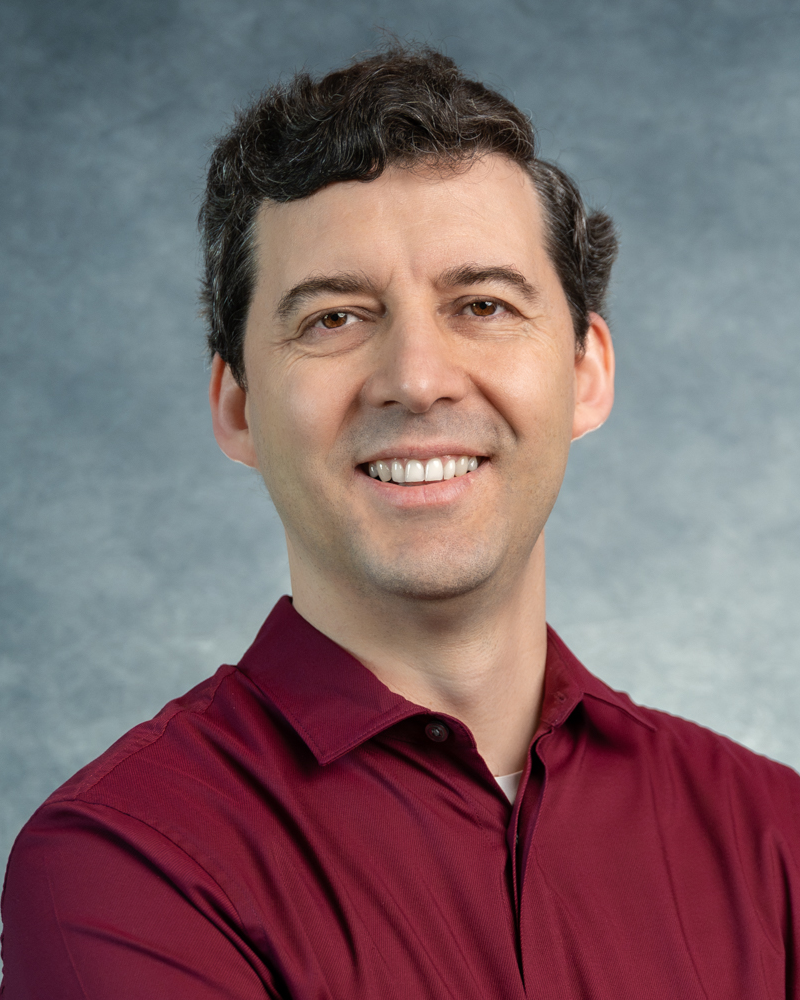}}]{Carlos Busso} 
(S'02-M'09-SM'13-F'23) is a Professor at Language Technologies Institute, Carnegie Mellon University, where he is also the director of the \emph{Multimodal Speech Processing} (MSP) Laboratory. His research interest is in human-centered multimodal machine intelligence and applications, focusing on the broad areas of speech processing, affective computing, multimodal behavior generative models, and foundational models for multimodal processing. 
He is an IEEE Fellow and an ISCA Fellow.
\end{IEEEbiography}

\vspace{-1.3cm}

\begin{IEEEbiography}[{\includegraphics[width=1in,height=1.25in,clip,keepaspectratio]{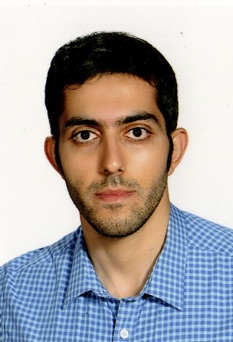}}]{Reza Lotfian} is a Senior Machine Learning Engineer at athenahealth, developing AI solution for healthcare industry. He earned a Ph.D. in Electrical Engineering from UT Dallas (2018), after an M.Sc. from Sharif University (2010) and a B.Sc. from Amirkabir University (2006). At UTD’s MSP Lab (2013–2018), he contributed to the development of the MSP-Podcast corpus. His interests include speech emotion recognition, affective computing, NLP, LLMs, and scalable ML systems.
\end{IEEEbiography}

\vspace{-1.3cm}

\begin{IEEEbiography}[{\includegraphics[width=1in,height=1.25in,clip,keepaspectratio]{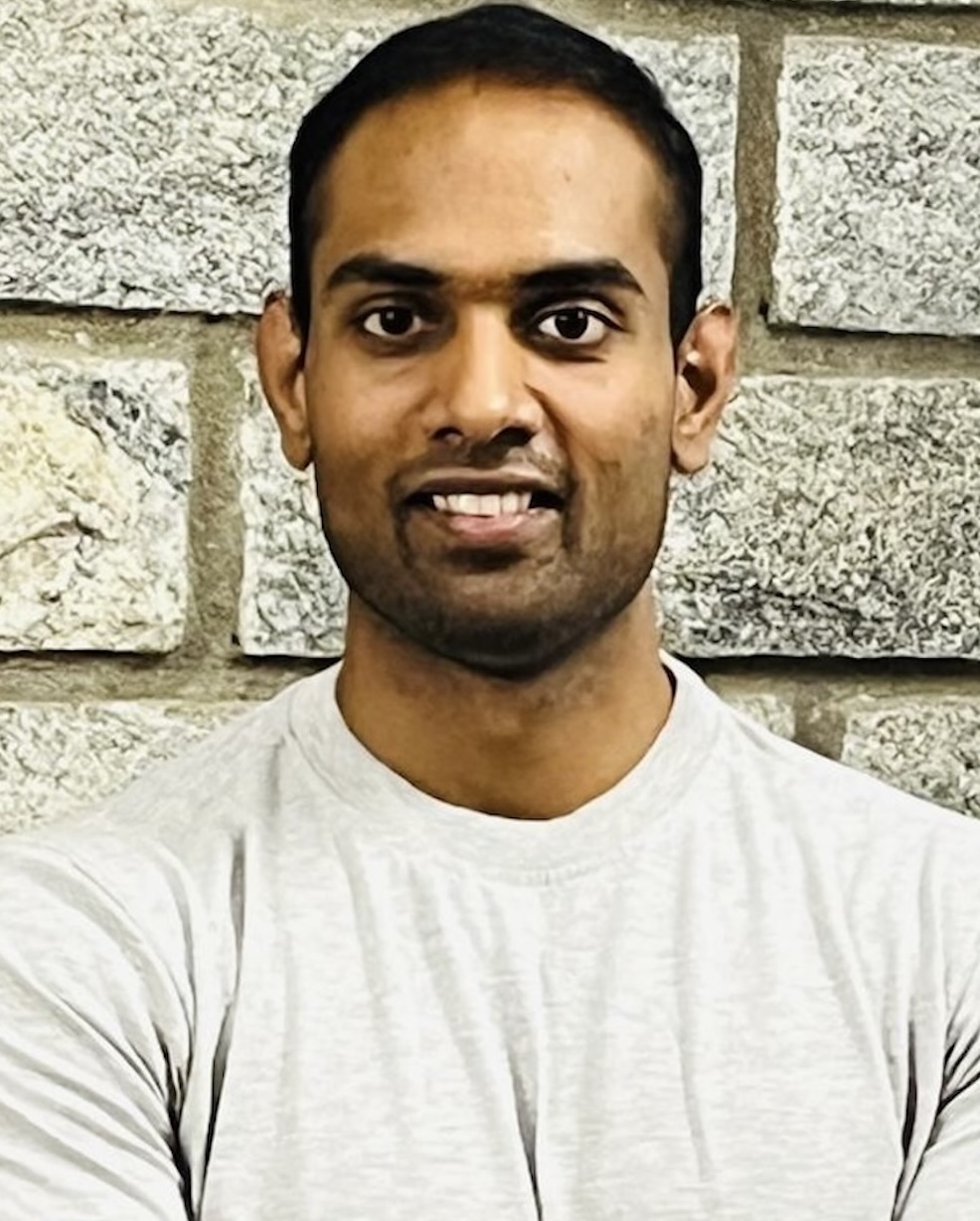}}]{Kusha Sridhar}
(Aug'21) received received his M.S. degree in electrical engineering from the University of Southern California (USC), Los Angeles, in 2015 and Ph.D. degree in electrical engineering from the University of Texas at Dallas, in 2021. He is currently a Sr. Manager at Accenture's Advanced Computational AI group. He has previously worked as a Staff Research scientist at Hippocratic AI Inc. His research interests include areas related to affective computing, conversational speech models and multi-modal signal processing.
\end{IEEEbiography}

\vspace{-1.3cm}

\begin{IEEEbiography}[{\includegraphics[width=1in,height=1.25in,clip,keepaspectratio]{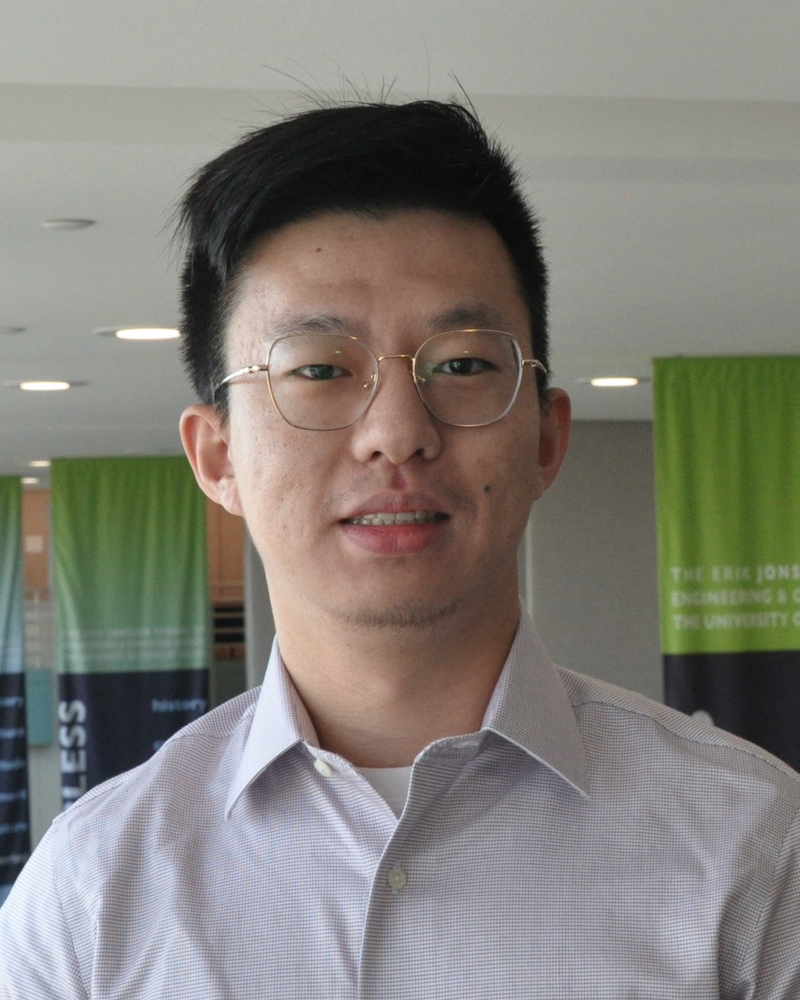}}]{Wei-Cheng Lin}
(S'16-M'23) received his PhD degree (2023) in electrical engineering from the University of Texas at Dallas (UTD). He is currently a research scientist at Bosch Research, Bosch Center for Artificial Intelligence, USA. His research focus on multimodal signal processing and deep learning. He was recognized with the Best Dissertation Award from the Association for the Advancement of Affective Computing (AAAC) in 2024.
\end{IEEEbiography}

\vspace{-1.3cm}

\begin{IEEEbiography}[{\includegraphics[width=1in,height=1.25in,clip,keepaspectratio]{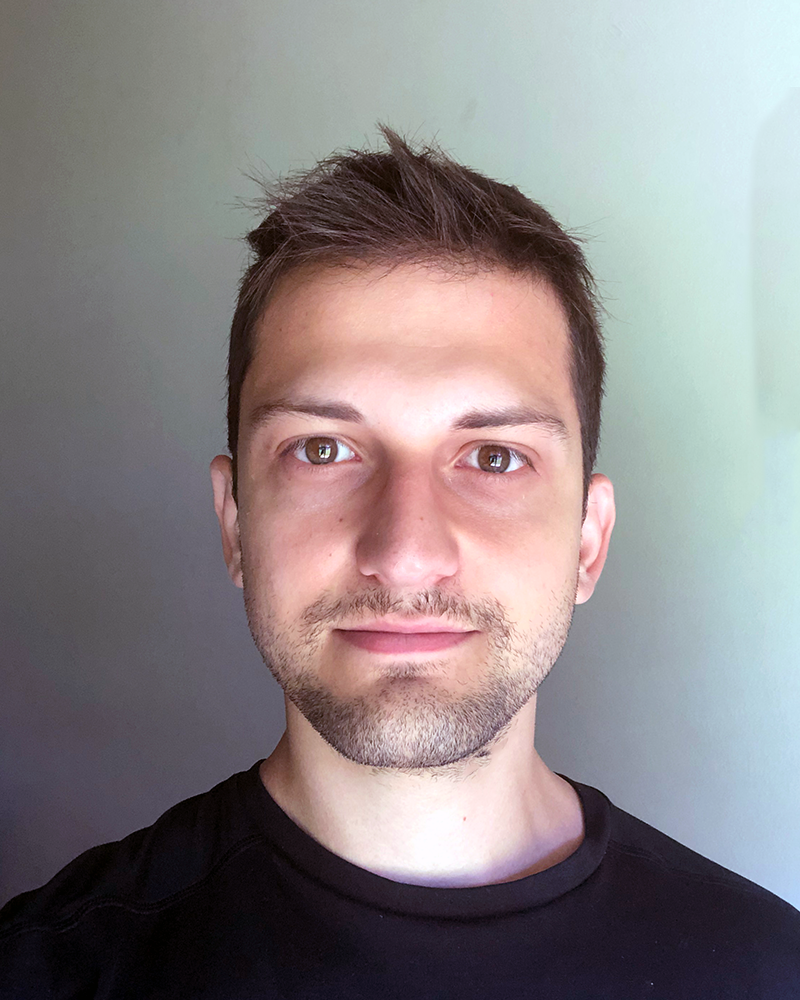}}]{Lucas Gon\c{c}alves}
(S'22–M'24) is an Applied Scientist at Amazon, USA. He received the Ph.D. degree in electrical engineering from The University of Texas at Dallas (UTD), Richardson, TX, USA, in 2024. From 2022 to 2024, he was a recipient of the Erik Jonsson School Excellence in Education Doctoral Fellowship. His research interests include multimodal signal processing and deep learning, with emphasis on audio–visual learning, speech and language technologies, and vision–language models.
\end{IEEEbiography}

\vspace{-1.3cm}

\begin{IEEEbiography}[{\includegraphics[width=1in,height=1.25in,clip,keepaspectratio]{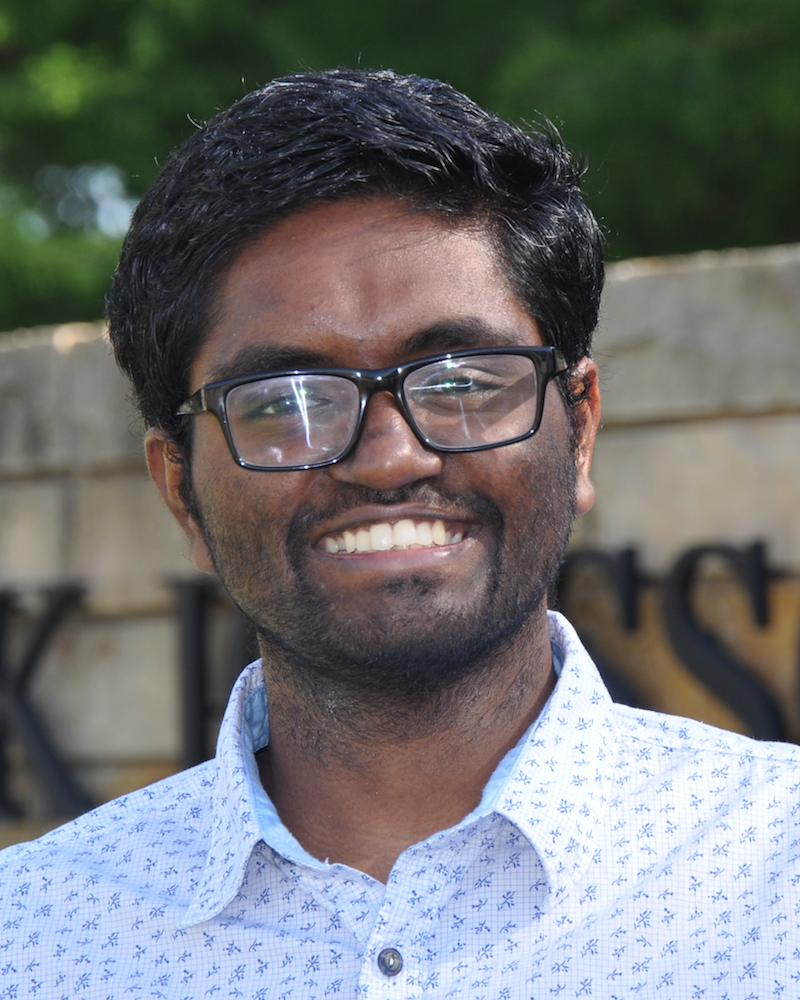}}]{Srinivas Parthasarathy}
(M'20) is a Senior Applied Scientist at Amazon. He received his Ph.D. degree in Electrical Engineering from The University of Texas at Dallas (UTD) in 2019. His research focuses on computer vision, multimodal large language models, multi-modal signal processing and affective computing. At UTD, he received the Ericsson Graduate Fellowship during 2013–2014. Previously, he has been a Research Intern with Amazon, Microsoft Research and Bosch Research and Training Center.
\end{IEEEbiography}

\vspace{-1.0cm}

\begin{IEEEbiography}[{\includegraphics[width=1in,height=1.25in,clip,keepaspectratio]{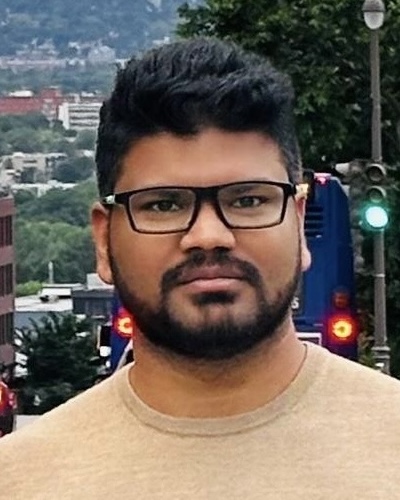}}]{Abinay Reddy Naini} (S'19)  is a PhD student in the Department of Electrical and Computer Engineering at the University of Texas at Dallas (UTD) and is currently working as a visiting researcher at the Language Technologies Institute, Carnegie Mellon University. He received his B.S. in Electrical Engineering from the National Institute of Technology, Warangal, India, and his M.S. in Electrical Engineering from the Indian Institute of Science (IISc). His research interests include affective computing, speech technology, and machine learning. 
\end{IEEEbiography}

\vspace{-1.0cm}

\begin{IEEEbiography}[{\includegraphics[width=1in,height=1.25in,clip,keepaspectratio]{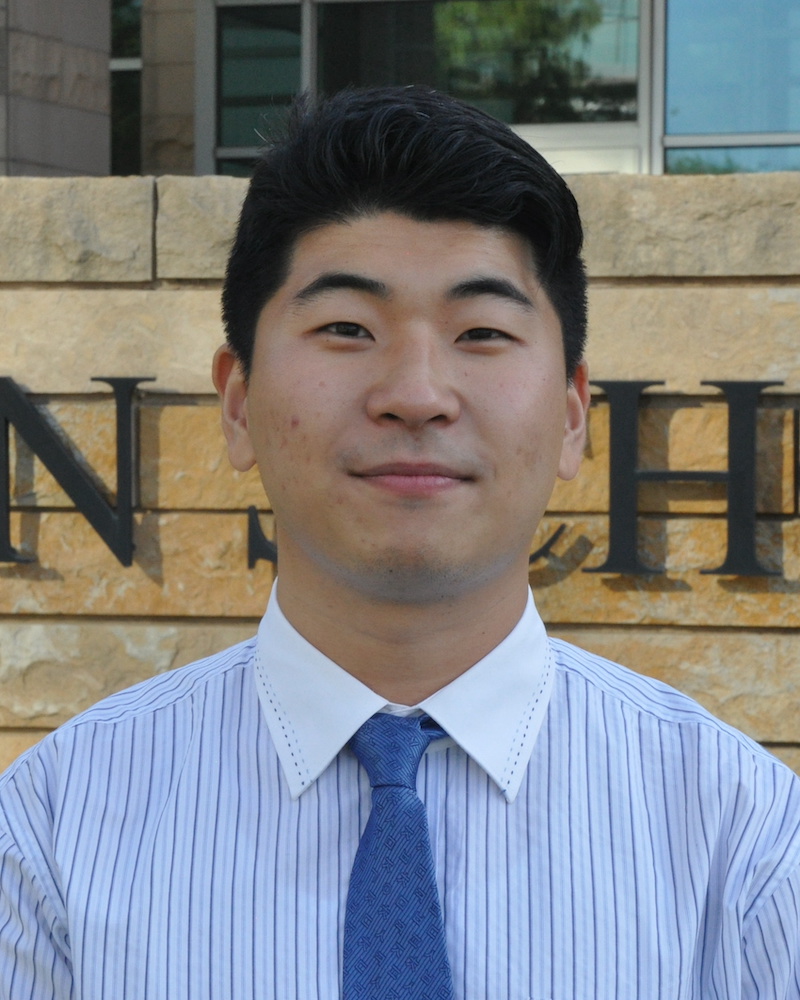}}]{Seong-Gyun Leem} is a research scientist in the Reality Labs at Meta Platforms, Inc. He received his B.S. and M.S. degrees in Computer Science and Engineering at Korea University, Seoul, South Korea, in 2018 and 2020, respectively. He received his Ph.D. degree in electrical engineering from the University of Texas at Dallas in 2024. His current research interests include speech synthesis, speech emotion recognition, noisy speech processing, and machine learning.
\end{IEEEbiography}

\vspace{-1.0cm}

\begin{IEEEbiography}[{\includegraphics[width=1in,height=1.25in,clip,keepaspectratio]{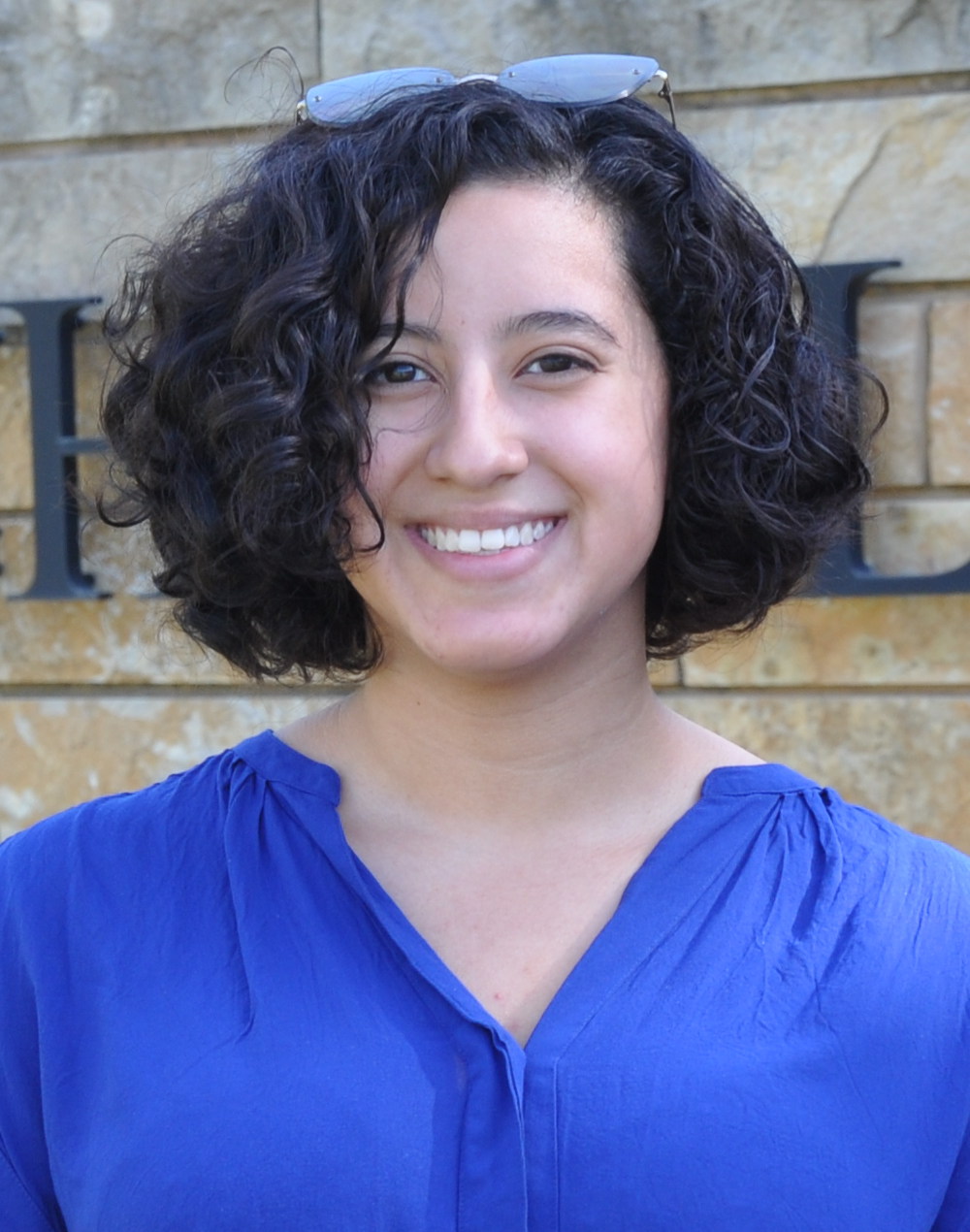}}]{Luz Martinez-Lucas} (S'21) is a PhD Student in the Electrical and Computer Engineering Department at the University of Texas at Dallas (UTD). She did her Bachelor's in Electrical Engineering at UTD. Her research interests include affective computing, speech technology, and machine learning. She is a student member of IEEE and AAAC.
\end{IEEEbiography}

\vspace{-1.0cm}

\begin{IEEEbiography}[{\includegraphics[width=1in,height=1.25in,clip,keepaspectratio]{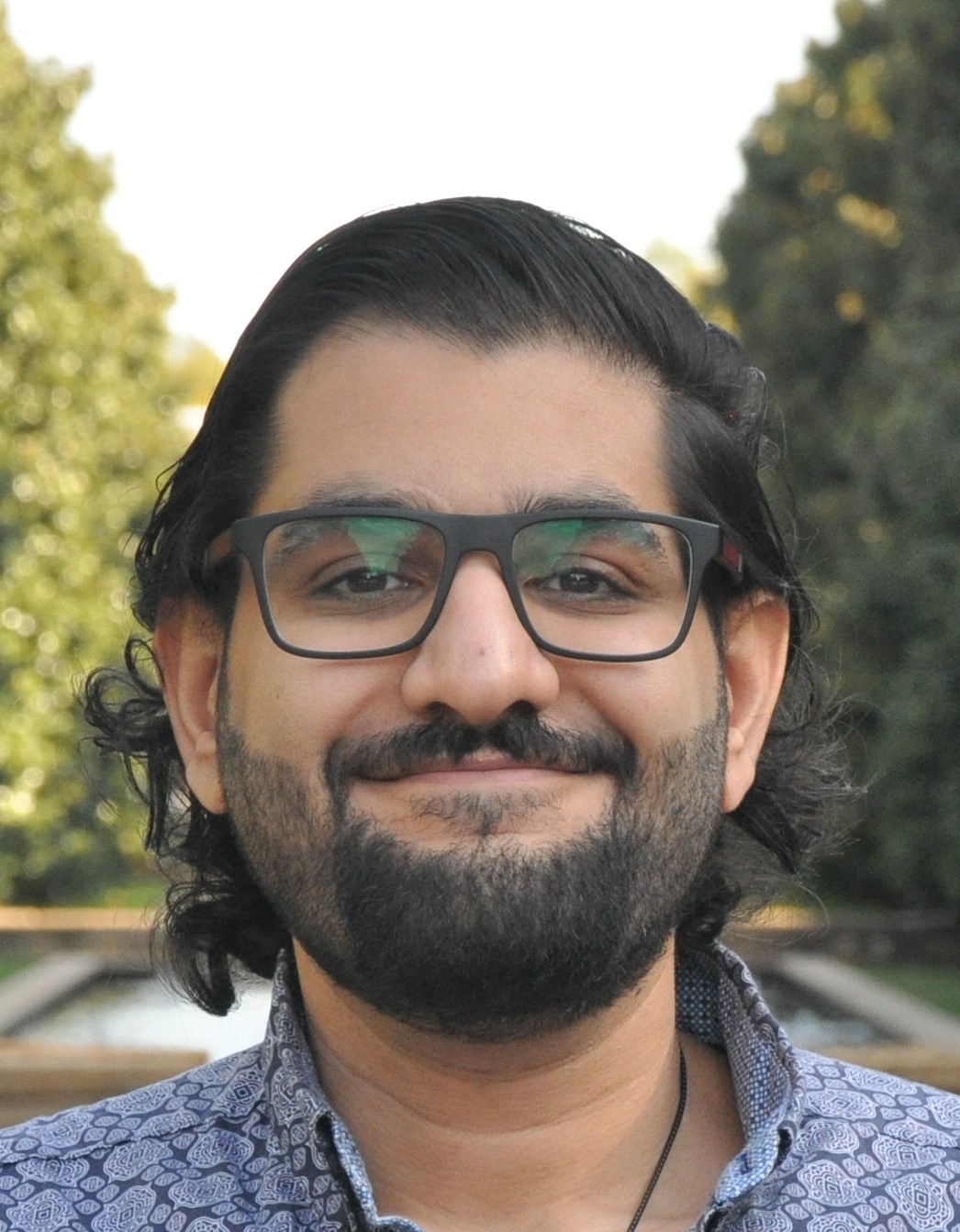}}]{Ali N. Salman} is a Research Scientist at ARRAY Innovation. He received his Ph.D. in Electrical Engineering from the University of Texas at Dallas in 2024, and his B.S. and M.S. degrees in Computer Science from Indiana State University in 2015 and 2017, respectively. His research interests include affective computing, retrieval-augmented generation (RAG) systems, and facial analysis.
\end{IEEEbiography}

\vspace{-1.0cm}

\begin{IEEEbiography}
[{\includegraphics[width=1in,height=1.25in,clip,keepaspectratio]{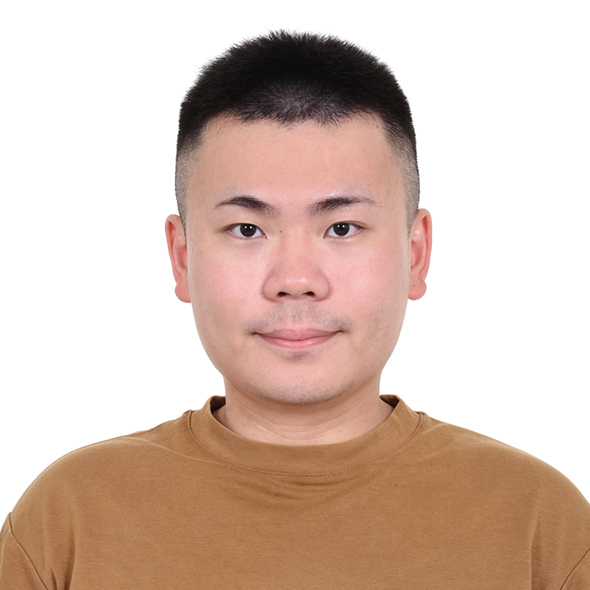}}]{Huang-Cheng Chou} (S'19–M'24) received the B.S. and Ph.D. degrees in electrical engineering from National Tsing Hua University (NTHU), Hsinchu, Taiwan, in 2016 and 2024, respectively. From 2021 to 2022, he was a Visiting Scholar at the Erik Jonsson School of Engineering and Computer Science, University of Texas at Dallas (UTD), Richardson, TX, USA. He is currently a Postdoctoral Scholar at the University of Southern California (USC). His research interests lie in affective computing.
\end{IEEEbiography}

\vspace{-1.0cm}

\begin{IEEEbiography}
[{\includegraphics[width=1in,height=1.25in,clip,keepaspectratio]{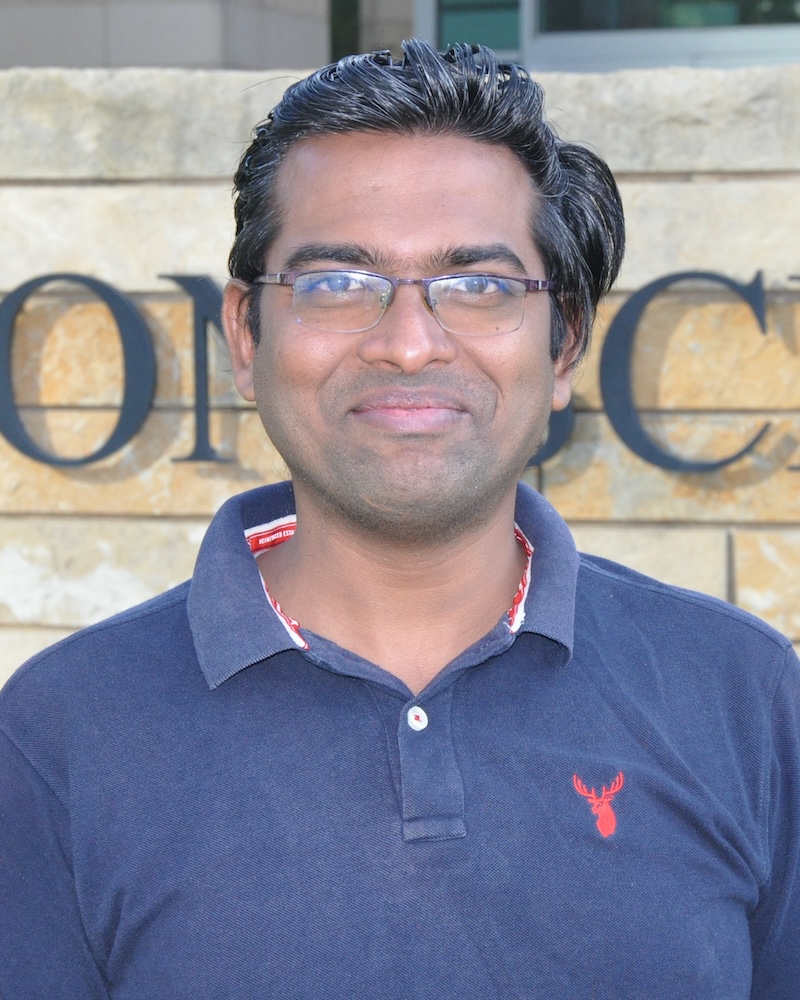}}]{Pravin Mote} is currently pursuing a Ph.D. in the Department of Electrical and Computer Engineering at the University of Texas at Dallas. He is also a visiting researcher at the Language Technologies Institute, Carnegie Mellon University. His research interests include speech technology, multimodal affective computing, and machine learning.
\end{IEEEbiography}




\end{document}